\documentclass[twocolumn,trackchanges]{aastex701}
\usepackage[english]{babel}
\usepackage{amsmath,amssymb}
\usepackage{graphicx}
\usepackage{subcaption, multirow, array}
\usepackage{bm}          
\usepackage[table, dvipsnames]{xcolor}      
\usepackage{booktabs}
\usepackage{hyperref}
\usepackage{colortbl}
\usepackage{tikz}
\usetikzlibrary{calc}

\newcommand{\taud}{\tau_{\rm decay}}
\newcommand{\taua}{\tau_{\rm A}}

\newcommand{\EQ}{\begin{equation}}
\newcommand{\EN}{\end{equation}}
\newcommand{\EQA}{\begin{eqnarray}}
\newcommand{\ENA}{\end{eqnarray}}

\newcommand{\Eq}[1]{Eq.~(\ref{#1})}
\newcommand{\Eqs}[2]{Eqs.~(\ref{#1}) and~(\ref{#2})}
\newcommand{\Fig}[1]{Fig.~\ref{#1}}

\newcommand{\Figs}[2]{Figs.~\ref{#1} and \ref{#2}}

\newcommand{\Tab}[1]{Table~\ref{#1}}

\newcommand{\BB}{\bm{B}}

\newcommand{\uu}{\bm{u}}

\newcommand{\AAA}{\bm{A}}
\newcommand{\JJ}{\bm{J}}
\newcommand{\nab}{\bm{\nabla}}

\begin{document}

\title{Spectral Evolution and Current Sheet Analysis as Probes of Reconnection-Mediated Decay in Magnetically Dominated  Turbulence}

\author[orcid=0000-0002-2969-1402]{Chandranathan Anandavijayan} 
\affiliation{International Center for Theoretical Sciences ICTS-TIFR, Bengaluru, India}
\email[show]{chandranathan.a@icts.res.in}

\author[]{Pallavi Bhat}
\affiliation{International Center for Theoretical Sciences ICTS-TIFR, Bengaluru, India}
\email[show]{pallavi.bhat@icts.res.in} 

\begin{abstract}
The decay of magnetically dominated turbulence exhibits robust inverse transfer of magnetic energy even in the absence of net magnetic helicity, challenging traditional cascade-based phenomenology. While recent studies suggest that magnetic reconnection governs the evolution of such systems, a comprehensive understanding has been lacking.
Here we test a reconnection-mediated model for decaying magnetic turbulence in two-dimensional (strict-2D), 2.5D, and three-dimensional (3D) systems with both helical and nonhelical initial conditions. We show that the magnetic-energy decay timescale scales with the Lundquist number in a manner consistent with Sweet–Parker–type reconnection rather than Alfvénic or purely resistive timescales. We develop a broken power-law model for the magnetic energy spectra and  provide analytic predictions for the temporal evolution of energy across both sub-inertial and inertial ranges, which are confirmed by high-resolution simulations. In nonhelical turbulence, these results favor anastrophy as the dominant constraint over helicity fluctuations.
Using Minkowski functionals to analyze reconnecting current sheets in real space, we find that the structures controlling the decay are substantially smaller than the global magnetic correlation scale, implying local Lundquist numbers well below the system-scale value. This explains the weak sensitivity of global decay laws to current-sheet resolution  and that the current-sheet aspect ratios converge toward Sweet–Parker predictions only at sufficiently high resolution.
Together, these results establish magnetic reconnection as the organizing principle underlying inverse transfer, spectral evolution, and decay in magnetically dominated turbulence, providing a unified picture applicable across dimensionality and helicity regimes with direct implications for astrophysical plasmas.
\end{abstract}

\keywords{ \uat{Primordial magnetic fields}{1294}, \uat{Magnetohydrodynamics}{1964}, \uat{Space plasmas}{1544}}

\section{Introduction}
Observation of TeV blazars through cosmic voids infer the presence of an intergalactic magnetic field of strength $\mathcal{O}\left(10^{-16}\,G\right)$ on $Mpc$ scales \citep{Neronov_Vovk_2010, Vovk_Korochkin_Neronov_Semikoz_2024}.  These observations establish a picture of a universe that is ubiquitously magnetised. Magnetic fields have been observed across a wide range of astrophysical systems from planets to galaxy clusters and mediums as dilute of the filaments in superclusters of galaxies \citep{Pignataro_O’Sullivan_Bonafede_Bernardi_Vazza_Carretti_2025}. In these environments, magnetic fields typically play an important role in their formation and evolution of structure. 

Although the initial seeding of magnetic fields remains uncertain, their subsequent amplification and maintenance are attributed to astrophysical processes such as dynamo action \citep{Brandenburg_Subramanian_2005}. Cosmic voids, however, are expected to be largely free of such activity \citep{Beck_Hanasz_Lesch_Remus_Stasyszyn_2013, Samui_Subramanian_Srianand_2018}
. The apparent presence of magnetic fields in these pristine regions, therefore, points to a primordial origin \citep{Brandenburg_Ghosh_Vazza_Neronov_2025}, with turbulent generation mechanisms operating during inflation or cosmological phase transitions. If confirmed by direct observation, such relic magnetic fields would influence a broad range of cosmological processes, including Big Bang nucleosynthesis, recombination \citep{Grasso_Rubinstein_2001, Jedamzik_Pogosian_2020, Schiff_Venumadhav_2025}, and the formation of large-scale structure \citep{Mtchedlidze_Domínguez-Fernández_Du_Brandenburg_Kahniashvili_O’Sullivan_Schmidt_Brüggen_2022}. Despite many theoretical works \citep{Subramanian_2016, Vachaspati_2021}, there is currently no consensus on how these primordial fields are generated in the early universe. In the absence of turbulent driving, their subsequent evolution is expected to be governed by decay dynamics of magnetohydrodynamic (MHD) turbulence.

The decay of three-dimensional MHD turbulence was previously understood as a forward-cascading process analogous to the purely hydrodynamic case, in which energy is transferred from large scales to small scales and dissipated. In the presence of net magnetic helicity, however, inverse transfer of magnetic energy to larger scales is expected due to the robust conservation of helicity in the limit of vanishing resistivity \citep{Christensson_Hindmarsh_Brandenburg_2001}.

A significant shift in this paradigm occurred with the numerical work of \citet{Brandenburg_Kahniashvili_Tevzadze_2015} and \citet{Zrake_2014}, who demonstrated that inverse transfer of magnetic energy can arise even in the absence of net helicity. Their simulations of freely decaying magnetically-dominated nonhelical turbulence showed robust growth of magnetic correlation lengths despite vanishing global helicity, challenging the notion that helicity conservation is a necessary condition for inverse energy transfer. While these studies established the phenomenon convincingly, the physical mechanism underlying non-helical inverse transfer remained unclear.

Subsequent work by \citet{Bhat_Zhou_Loureiro_2021} identified magnetic reconnection as the key organizing principle governing this behavior in 3D. It was shown that reconnection-mediated restructuring dominate the dynamics in energy transfers, leading naturally to inverse transfer and growth of magnetic coherence scales. An important finding was that the decay time scale is related to reconnection as opposed to Alfvénic timescale invoked by previous models of decaying non-helical MHD turbulence. This was confirmed by \citet{Brandenburg_Neronov_Vazza_2024} who explicitly show that the decay timescale depends on resistivity, lending strong support to interpretations in which reconnection  \citep{Bhat_Zhou_Loureiro_2021}, rather than purely Alfvénic transport, controls the evolution of magnetically dominated turbulence. 
Complementary work by \cite{Hosking_Schekochihin_2021} further clarified the phenomenology of decaying magnetically dominated turbulence, reinforcing the role of reconnection. 

Besides the characteristic timescale, the evolution of the system is further constrained by the presence of (approximately) conserved quantities. Two such candidates have been proposed in recent work: (i) anastrophy \citep{Bhat_Zhou_Loureiro_2021}, and (ii) helicity fluctuations \citep{Hosking_Schekochihin_2021}. While anastrophy, given by $\int A^2 d\bm{x}$ with $\bm{A}$ being the vector potential, is ideally conserved only in two dimensions, it has been shown to exhibit remarkably robust conservation properties even in three-dimensional, magnetically dominated decaying turbulence, as long as it is nonhelical. This behavior has been attributed to the quasi-dimensionalization of the flow \citep{Dwivedi2024}. Under the assumption of anastrophy conservation, a power-law decay of the magnetic energy, $E \sim t^{-p}$, yields a decay exponent $p=1$.

An alternative proposal based on the conservation of helicity fluctuations, quantified by the integral $I_H  = \int \, d^3r\, \langle h(\mathbf{x})h(\mathbf{x}+\mathbf{r})\rangle$, where $h(x)=\bm{A}\cdot{\bm B}$, leads to a slightly steeper decay exponent, $p \simeq 1.18$ \citep{Hosking_Schekochihin_2021}. This quantity has also been shown to be approximately conserved in nonhelical decaying systems \citep{Zhou_Sharma_Brandenburg_2022}. However, direct numerical comparisons indicate that anastrophy is more robustly conserved, particularly in the low resistivity limit  \citep{Dwivedi2024}.
More generally, turbulent systems may admit multiple ideal or quasi-ideal invariants. Identifying which of these plays the dominant role in constraining the dynamics and thereby controlling the decay laws remains a nontrivial and system-dependent question.

In this paper, we establish the role of magnetic reconnection in three-dimensional, magnetically dominated decaying turbulence and the associated conserved quantities. To this end, we first introduce the Kolmogorov model for decaying turbulence and its analytical solution in Section~\ref{sec:Model for decaying turb}. We then test our theoretical predictions using three-dimensional numerical simulations, the details of which are described in Section~\ref{sec:numerical setup}. In Section~\ref{sec:timescale}, we perform an analysis similar to that of \citet{Brandenburg_Neronov_Vazza_2024} to identify the characteristic decay timescale. To further support our theory for the decay timescale and the associated conserved quantities, we propose a broken power-law model for the evolution of various spectral modes and numerically verify our predictions in Section~\ref{sec:specevol}. Finally, a detailed analysis of local current-sheet structures and reconnection scaling is presented in Section~\ref{sec:recres}.
\section{Model for decaying turbulence\label{sec:Model for decaying turb}}

The current paradigm in understanding decaying turbulence is via \citet{Kolmogorov41b}, which proposes that the decay is modelled as a power law in time. The model is a simplified one which describes the evolution of the total energy and the correlation scale.
This implicitly assumes that the spectrum is steeply peaked one with the inertial range occurring after the peak set by local direct cascade. 

The decay is modelled by the standard power law differential equation, 
\begin{equation}
    \label{modelpart1}
    \frac{d E_M}{dt} = - \frac{E_M}{\taud (E_M,\xi_M)}, 
\end{equation}
along with 
\begin{equation}
    \label{modelpart2}
E_M^{\alpha/2}\xi_M\sim \text{const.}, 
\end{equation}
where $\xi_M$, the characteristic scale of the system, is taken to be the correlation scale of the magnetic field. 
The model is constrained by specifying two ingredients, (i) the form of $\tau_{decay}(E_M,\xi_M)$ and (ii) the conserved quantity in the system which describes the relation between $E_M$ and $\xi_M$ given in \Eq{modelpart2}. For example, $\alpha=2$ in \Eq{modelpart2} when the governing conserved quantity is magnetic helicity, $\int \AAA\cdot \BB~dV$. 

\subsection{Decay timescale}
In the magnetically dominated case, we assume that the time scale is a function of Lundquist number $S = \sqrt{2E_M}\,\xi_M/\eta$ and Alfven timescale $\tau_A=\xi_M/\sqrt{2 E_M}$. 
\begin{equation}
    \label{decay timesscale in terms of S}
    \taud = S^n \tau_A,
\end{equation}
where $n=0$ and $n=1$ pertains to Alfv\'en timescale and resistive diffusion timescale respectively. In particular, recently \cite{Bhat_Zhou_Loureiro_2021,Hosking_Schekochihin_2021,Dwivedi2024} proposed that the decay in 3D decaying turbulence is governed by the reconnection time scale. In the case of lower values of $S$, the relevant model that can describe reconnections in such a quasi-steady system is the Sweet-Parker (SP) model, where $n=1/2$. 

\subsection{Conserved quantity}

Magnetic turbulence can be categorized based on the presence of magnetic helicity in the system. Based on the `strength' of helicity in the system, we can have (i) fully helical, (ii) fully nonhelical, and (iii) partially helical cases. 
The strength of helicity is quantified as follows. 

Suppose $H(k)$ and $M(k)$ are the one-dimensional shell-averaged helicity and energy spectrum. 
For the fully helical case, the total magnetic energy $E_M$ is equal to that calculated using the helicity \textit{i.e.} $\int ~k~H(k)dk= 2\int M(k) dk = 2E_M$, as expected from the realizability condition $\vert kH(k)\vert \le 2M(k)$ \citep{Frisch_Pouquet_LÉOrat_Mazure_1975}.

In the nonhelical case, the total magnetic helicity, $\int \AAA\cdot\BB~dV = \int H(k)dk$, is zero. And partially helical is when the fraction of helical energy to the total magnetic energy, $\int (H(k)/2)dk/ \int M(k)dk$, is between zero and one.
We will focus only on the first two of the three cases in this paper. 

While it is obvious that the fully helical case is constrained by the magnetic helicity with $\alpha=2$, it is not as manifestly clear for the nonhelical case. This is because, in the helical case, magnetic helicity is an ideal invariant. However, such an ideal invariant, which could control the dynamics, was found to be lacking in the nonhelical case. Even though \citet{Brandenburg_Kahniashvili_Tevzadze_2015} had explored the conservation of particular components of the vector potential at the level of temporal scalings, a firm theoretical understanding was found lacking.

\citet{Bhat_Zhou_Loureiro_2021} proposed anastrophy as a relevant candidate based on quasi-two-dimensionalization of the 3D nonhelical turbulence.
An alternate proposal claimed that helicity fluctuations can constrain this system \citep{Hosking_Schekochihin_2021}, given by the integral $I_H$, which leads to $\alpha=4/5$ in \Eq{modelpart2}. 
Subsequent work by \cite{Dwivedi2024} firmly established the robust conservation of anastrophy due to quasi-two-dimensionalization while critically analyzing $I_H$. They showed that the system lacks sufficiently strong local helical patches, as required by $I_H$ to be able to constrain the system effectively. 

\subsection{Model solutions}
The knowledge of the decay timescale and the conserved quantity can be encapsulated by two variables, $\{n,\alpha \}$. To obtain the scaling laws in time \textit{i.e.} $E_M \sim t^{-p}$ and $\xi_M \sim t^{q}$, in terms of these variables, we solve \Eq{modelpart1} using the initial state, 
\EQ
E_M(t) = \frac{E_M(0)}{\left(1 + C \,t\right)^{p(n,\alpha)}},  
\label{soln}
\EN
where the temporal exponents are given as a function of $n$ and $\alpha$,
\begin{equation}
    \label{scaling exponents}
    \begin{split}
        p(n,\alpha) &= \frac{2}{\alpha+1 + n\,(\alpha-1)},\\
        q(n,\alpha) &= \frac{\alpha}{\alpha+1 + n\,(\alpha-1)}.
    \end{split}
\end{equation}
Henceforth, we will drop the arguments from $p$ and $q$ for ease of notation. 

In the literature, these exponents have been routinely reported for both helical and nonhelical decaying MHD systems in the last decade or so. However, the fundamental debate regarding the time scales and conserved quantities remains. As a step towards clarifying this conundrum, we extend this philosophy beyond $r.m.s.$ quantities to the evolution of spectral modes in section~\ref{sec:specevol}.
But first, we turn towards probing $\taud$ in our direct numerical simulations. 
\section{Numerical Setup\label{sec:numerical setup}}
We solve incompressible, isothermal MHD equations in the Weyl gauge using the \textsc{Pencil code} \citep{Pencilcode},

\begin{eqnarray}
\frac{\partial~\mathrm{ln}~\rho}{\partial t} + \uu\cdot\nab~\mathrm{ln}~\rho&=& - \nab \cdot \uu, \\
\frac{\partial \uu}{\partial t}+\uu\cdot\nab\uu &=& -c_s^2\nab~\mathrm{ln}~\rho + \frac{\JJ\times \BB}{\rho} + \frac{\bm{F}_{visc}}{\rho},\\
\frac{\partial\AAA}{\partial t} &=& \uu \times \BB - \eta \mu_0 \JJ,
\end{eqnarray}
where $\uu$ is the fluid velocity field, $\BB = \nab \times \AAA$ is the magnetic field, with $\AAA$ the magnetic vector potential, and $\JJ = \nabla \times \BB/\mu_0$ is the current density, with $\mu_0$ the vacuum permeability, $\rho$ is the mass density, and $\eta$ is the magnetic diffusivity. The scalar potential $\Phi = 0$ (Weyl gauge), and the viscous force $\bm{F}_{visc} = \nab \cdot 2\nu \rho \bm{S}$, where $\nu $ is the kinematic viscosity, and $\bm{S}$ is traceless rate of strain tensor with components $\bm{S}_{ij} = \frac{1}{2} (u_{i,j} + u_{j,i}) - \frac{1}{3} \delta_{ij}\nab \cdot \uu$ (commas denote partial derivatives).

The outputs of the code are nondimensionalized in terms of the system size $L_S$, isothermal sound speed $c_s$, initial density $\rho_0$, and $\sqrt{\mu_0\, \rho_0\, c_s^2}$. In particular, the system size $L_S$ is taken to be $2\pi$, so that the wavenumbers are positive integers.

We perform 2D, 2.5D, and 3D simulations with various resolutions and diffusivities as given in \Tab{tab:sim_params}. All these simulations are performed in a periodic domain with initial uniform density, zero velocity, and Gaussian random magnetic field. The initial magnetic field energy spectrum has a $k^4$ dependence on wavenumber $k$ peaking at a particular $k_p$ and falling afterwards. The initial r.m.s. magnetic field $B_{\rm rms,0}$ is set close to $0.4$. 

\begin{table}[h!]
\caption{Simulation parameters for 2D, 2.5D, and 3D runs. $S$ is the initial Lundquist number.}
\label{tab:sim_params}
\centering
\begin{tabular}{|c|c|c|c|c|c|}
\hline\hline
Dim & Helicity & Resolution & $\nu=\eta$ & $k_{p,0}$ & $S/10^3$ \\
\hline\hline
\multirow{6}{*}{\shortstack{2D\\ \& \\2.5D}}
 & \multirow{6}{*}{\shortstack{Hel\\ \& \\ Nonhel}}
 & 256  & 0.4  & 60 & 1.1 \\
 &  & 512  & 0.4  & 60 & 1.1 \\
 &  & 1024 & 0.4  & 60 & 1.1 \\
 &  & 2048 & 0.4  & 60 & 1.1 \\
 &  & 2048 & 0.2  & 60 & 2.2 \\
 &  & 2048 & 0.1  & 60 & 4.3 \\
\hline
\multirow{3}{*}{3D}
 & \multirow{3}{*}{Nonhel}
 & 1024 & 1.3  & 25 & 0.8 \\
 &  & 1024 & 0.65  & 25 & 1.7 \\
 &  & 1024 & 0.325 & 25 & 3.3 \\
\hline
\end{tabular}
\end{table}
\section{Decay timescale in simulations\label{sec:timescale}}
To determine the characteristic decay timescale of magnetic energy, we begin with an approach similar to that used by \citet{Brandenburg_Neronov_Vazza_2024}, but with a different method for calculations.
We define a dimensionless decay coefficient,
    \EQ
        C_M = \frac{\taud}{\taua} = \left(-\frac{d~\ln E_M}{dt}\right)^{-1} \, \frac{v_A}{\xi_M} . 
        \label{cm}
    \EN

To evaluate the quantity $\left(- d \ln E_M / dt\right)^{-1}$, we consider two options. One option is to compute the derivative directly from the numerical time series $E_M(t)$. Alternatively, and more robustly, we can fit $E_M(t)$ with an appropriate functional form and evaluate the derivative analytically from the fit. We adopt the latter option.

Specifically, we use the functional form given in \Eq{soln}, which is the solution of the model equations \Eqs{modelpart1}{modelpart2} subject to the appropriate initial conditions. For convenience, we reproduce it here:
\EQ
E_M(t) = \frac{E_M(0)}{\left(1 + C \,t\right)^p},
\label{fit}
\EN
where, $C$ and $p$ are fit parameters obtained from the simulation data.

The Alfvén speed $v_A$ appearing in \Eq{cm} is computed directly from the simulation data. Similarly, the characteristic scale $\xi_M$ may be identified with the magnetic integral scale, defined from the magnetic energy spectrum as, 
\EQ
\xi_M = \frac{\int (2\pi/k)~M(k) dk}{E_M}. 
\EN
However, we find that $L_{\rm int}$ does not reliably reproduce the expected behaviour of the characteristic scale $\xi_M$ implied by \Eq{modelpart2}. This discrepancy arises from the discrete evolution of the peak spectral mode in numerical simulations, which limits the fidelity of scale-based diagnostics derived from the spectrum. As a result, the integral scale does not always track the theoretical scaling behaviour accurately (see Appendix \ref{sec:evol of lint}).

\begin{figure*}
    \centering
    \includegraphics[width=0.3\linewidth]{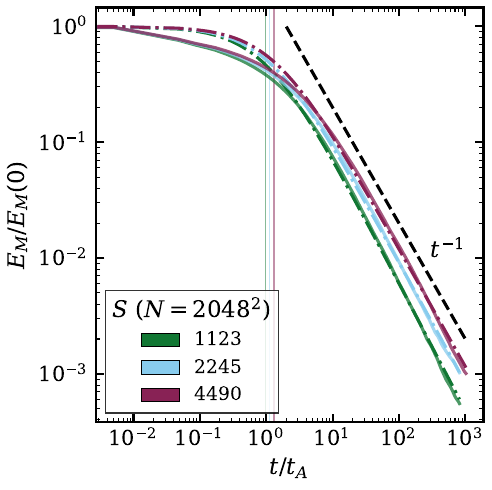}
    \includegraphics[width=0.3\linewidth]{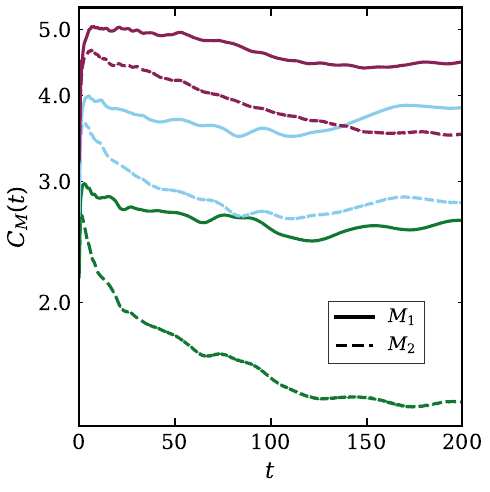}
    \includegraphics[width=0.3\linewidth]{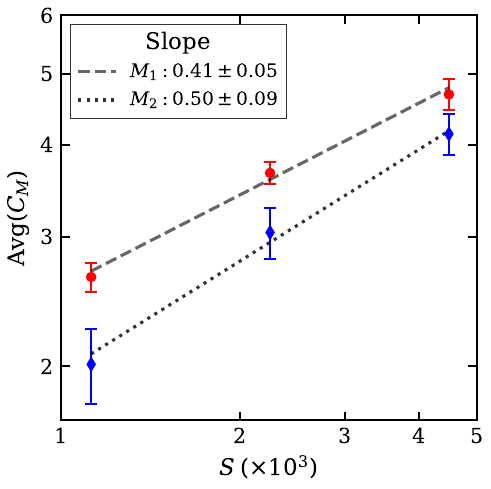}
    \centering
    \includegraphics[width=0.3\linewidth]{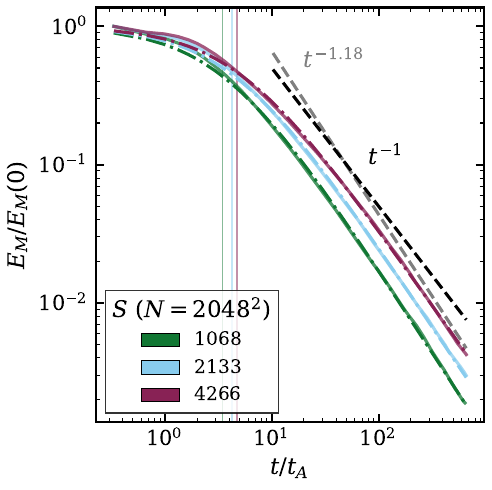}
    \includegraphics[width=0.3\linewidth]{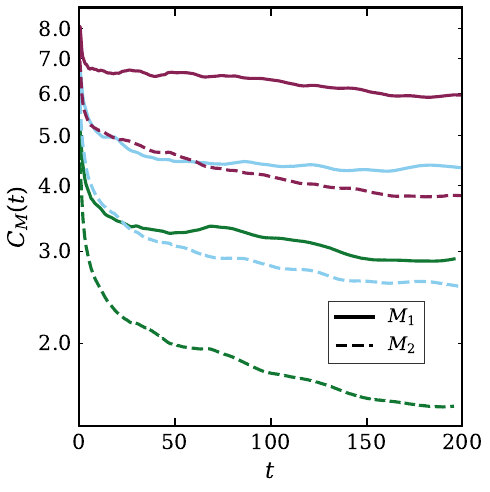}
    \includegraphics[width=0.3\linewidth]{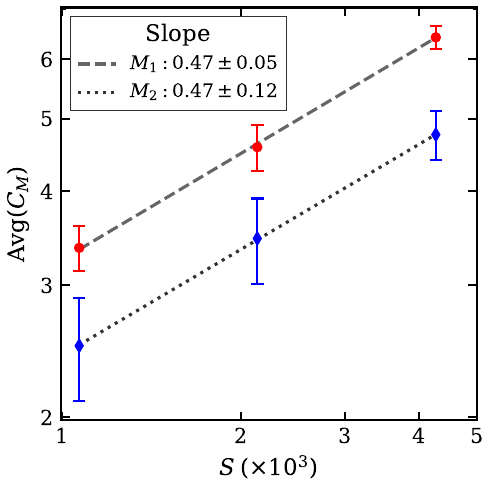}
    \centering
    \includegraphics[width=0.3\linewidth]{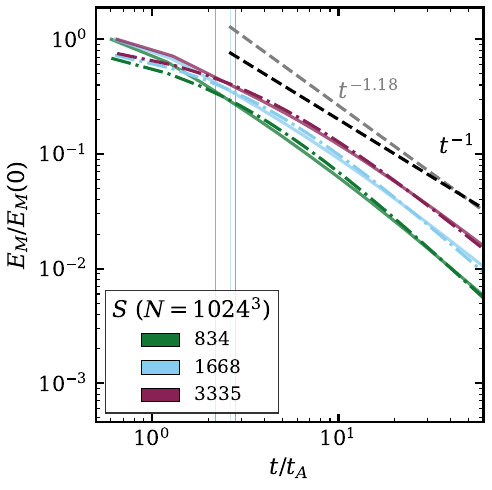}
    \includegraphics[width=0.3\linewidth]{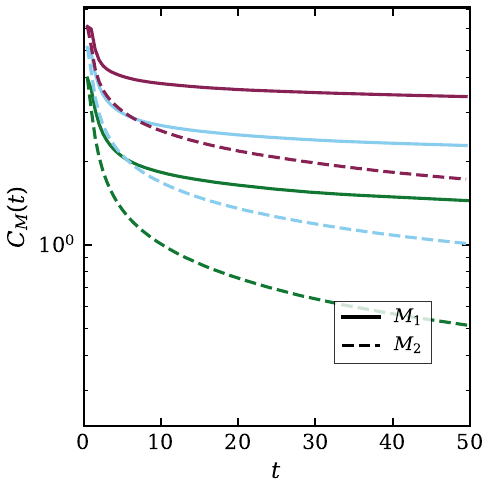}
    \includegraphics[width=0.3\linewidth]{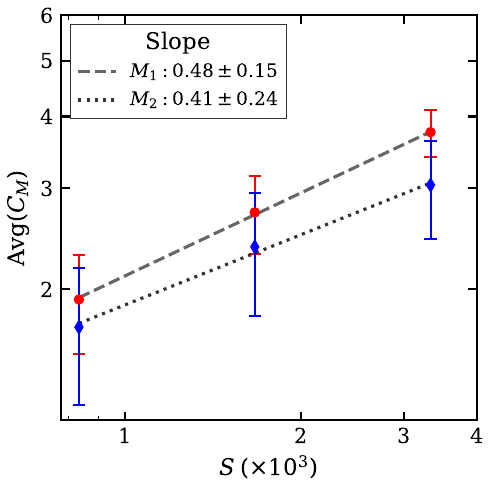}
   \caption{\textbf{Scaling of the decay time with the Lundquist number:} From top to bottom, rows correspond to strict 2D , 2.5D nonhelical, and 3D nonhelical simulations. The \textbf{left} column shows the temporal evolution of magnetic energy; dash--dotted curves indicate fits using Eq.~(\ref{fit}), and vertical solid lines mark the corresponding $1/C$. The \textbf{middle} column shows the evolution of $C_M(t)$, defined in Eq.~(\ref{cm}), for simulations with increasing initial Lundquist numbers $S$. The \textbf{right} column shows the average $C_M$ as a function of the initial Lundquist number. Best-fit scalings and associated uncertainties are indicated in the legends and are consistent with SP--type reconnection.
  }
    \label{fig:Cm_fit_t_S}
\end{figure*}

To address this issue, we also adopt an alternative approach in which we directly assume the scaling relation from \Eq{modelpart2}, $\xi_M \sim E_M^{-\alpha/2}$. Since our primary goal in this part of the analysis is to constrain the decay-time related Lundquist number exponent $n$, rather than independently determining both $n$ and $\alpha$, this assumption allows us to proceed without introducing additional uncertainty from scale measurements. 

In summary, both approaches use the fitted form \Eq{fit} to obtain $\left(- d \ln E_M / dt\right)^{-1}$. In method-1 ($M_1$), the Alfvén speed $v_A$ and characteristic scale $\xi_M$ are computed directly from the simulation data. In method-2 ($M_2$), $v_A$ is derived from the data and $\xi_M$ is obtained from the model relation \Eq{modelpart2}.

For comparison, in a previous study by \cite{Brandenburg_Neronov_Vazza_2024}, the quantity $C_M$ defined in \Eq{cm} was also evaluated, but using a simpler decay model $E_M(t) \sim t^{-p}$. In that work, the energy evolution was not explicitly fitted, and only an approach equivalent to our method-1 was employed. At $t \gg 1/C$, we expect our results from method-1 to be consistent with those of \cite{Brandenburg_Neronov_Vazza_2024}.

Finally, we use the calculated $C_M(t)$ to plot an averaged $C_M$ as a function of the initial Lundquist number $S$ and extract the exponent $n$. We use the initial $S$, rather than the Lundquist number calculated during the simulation, for the following reason. At low Lundquist numbers, the temporal decay exponent of the magnetic energy $E_M$ is observed to be larger than unity, approaching the expected value as the resistivity is reduced. Similarly, the correlation scale $\xi_M=L_{\rm int}$ grows with time with an exponent that is typically smaller than $0.5$
. Since the Lundquist number is constructed from the product of these quantities, it inherits the largest deviations present in each. In contrast, the quantity $C_M$, which involves the ratio of these measures, benefits from partial cancellation of low--Lundquist-number deviations, making it a more robust.

We present results for all three geometries: 2D, 2.5D, and 3D nonhelical turbulence, in \Fig{fig:Cm_fit_t_S}, shown in the top, middle, and bottom rows, respectively.
In the leftmost panels, we demonstrate that the functional form given by \Eq{fit} provides an excellent fit to the magnetic energy evolution $E_M(t)$ once the initial transient phase has passed. In all cases, the fitted curves closely track the simulation data over the subsequent decay phase, with reduced $\chi^2$ values close to unity, indicating statistically robust fits.

According to our model, the dimensionless decay coefficient scales with the Lundquist number as $C_M \propto S^n$. To examine this, we plot $C_M(t)$ in the middle panels of \Fig{fig:Cm_fit_t_S} using both method-1 and 2. We find that, after the transient regime and up to the time beyond which either resistive effects or box-scale effects could become more pronounced, $C_M(t)$ remains approximately constant, more clearly in the case of method-1 marked by solid curves. This weak time dependence is already suggestive of a scaling exponent $\alpha$ close to unity. 

Moreover, this near constancy of $C_M$ becomes increasingly pronounced as the Lundquist number $S$ is increased. At higher $S$, resistive effects are further suppressed, allowing the system to more closely approach the ideal limit and improving the conservation properties of the relevant invariant. 
Taken together, these results lend support to anastrophy, as the dynamically relevant conserved quantity controlling the decay, consistent with $\alpha = 1$.

In the right-most panels of ~\Fig{fig:Cm_fit_t_S}, we find that the average $n$ estimated is around $\sim 0.41$, $0.47$, and $0.48$ for strict 2D, 2.5D, and 3D cases respectively from method-1.
However, from method-2, the trend is reversed and we obtain $n \sim 0.5$, $0.47$, and $0.41$ for strict 2D, 2.5D, and 3D cases, respectively, more consistent with theoretical expectations.
This clearly indicates that the decay is not happening on Alfv\'enic time scales or the diffusive time scale.
The value of $n$ is somewhat smaller than that expected for SP reconnection, $n=0.5$, but it is consistent with it within the error bars in 2.5D and 3D cases. 
To calculate the error bars, we consider $C_M(t)$ over a specified time window, from $t = 1\,\tau_A$ up to the time when the correlation length reaches approximately one-fifth of the box size. This window is divided into equidistant bins, and within each bin, we calculate the mean value of $C_M(t)$. The overall mean across all bins provides an estimate of $C_M$, while the standard deviation of the bin means gives an error that is reflective of the weak time variability of $C_M$.

An interesting observation is that the three vertical lines marked in the leftmost panels in the \Fig{fig:Cm_fit_t_S} correspond to the timescale $1/C$ derived from the fit in \Eq{fit}. This visually seems to mark the knee in the curves or the transient timescale and that coincides with the Alfv\'enic timescale. Thus, our model in \Eq{fit}, doesn't really account for this initial transient, but we are anyway interested in the late-time (compared to Alfvenic timescale) dynamics which it models reasonably well. 

A similar exercise in the helical run cannot be performed because the Lundquist number evolves with time, $S\sim t^{2/7}$. Hence, we could not cleanly isolate $n$ in the helical cases.

\section{Model for the Spectral Evolution\label{sec:specevol}}

In decaying magnetic turbulence, inverse transfer is observed in both two and three dimensions, in contrast to the purely hydrodynamic case, where inverse transfer occurs only in two dimensions \citep{Brandenburg_Kahniashvili_2017}. Inverse transfer of energy implies a growth of energy in scales above the inertial range over time. Consequently, the temporal evolution of spectral energy in the sub-inertial and inertial ranges is expected to differ. We therefore propose a model that analytically tracks the evolution of energy in these distinct spectral regimes.


We first note that the magnetohydrodynamic equations are  scale invariant under simultaneous rescalings of space, time, velocities, and dissipation coefficients.
\cite{Olesen97} had applied dimensional arguments to obtain a self-similar form for the evolution of energy. However, this method is inadequate because the temporal scalings are determined by the initial spectrum alone. To remedy this, \citet{Brandenburg_Kahniashvili_2017} proposed an alternate self-similarity form, in terms of $\xi(t)$ and $\kappa = k~\xi(t)/2\pi$,
\begin{equation}
    \label{self sim tina}
    M(\xi,\kappa) = \xi^{-\beta}\phi(\kappa),
\end{equation}
where $\beta = p/q-1$. Here, $\xi$ can be thought of as $\xi_M$ and $k$ is the wavenumber of one-dimensional spectrum reduced from 3D. 

Previous studies probed the magnetic spectral evolution in only the sub-inertial range in the limit $\kappa \to 0$ \citep{Brandenburg_Sharma_Vachaspati_2023}. Here, we extend the analysis to the full spectrum by proposing a model of broken power-law form spanning both the sub-inertial and inertial ranges, matched at the peak wavenumber or equivalently, the inverse of the integral scale, given by 


\begin{equation}
    \label{self similar function}
    \phi(\kappa) \sim \begin{cases}
     \kappa^r , &\kappa < 1  \\
    \kappa^{-f}, &\kappa\geq 1
    \end{cases},
\end{equation}
where $r$ and $f$ are specified for the rising (sub-inertial) and falling (inertial) part of the spectrum, respectively. We rewrite $M$ in $\{k,t\}$ variables to obtain the time evolution of various spectral modes, 
\begin{equation}
    \label{eqn:time evol of energy}
    M(k,t) 
    \sim \begin{cases}
        k^r t^{q(-\beta+r)}, & k<k_p(t)\\
        k^{-f} t^{q(-f-\beta)}, & k\geq k_p(t).
    \end{cases}
\end{equation}


Written in terms of $\{n,\alpha\}$ using \Eq{scaling exponents}, the temporal exponents for spectral modes are,
\begin{equation}
\label{eqn: defn of sf sr}
\begin{split}
    \sigma_r(n,\alpha) &:= q(-\beta+s) = \frac{-2 + \alpha\, (1+s)}{\alpha+1 + n\,(\alpha -1)}, \\
    \sigma_f(n,\alpha) &:= q(-\beta-r) = \frac{-2 + \alpha\, (1-r)}{\alpha+1 + n\,(\alpha -1)} .
\end{split}
\end{equation}


In our previous studies, we have found that the nonhelical and helical energy spectra typically develop inertial range slopes of $k^{-2}$ and $k^{-2.5}$, respectively, corresponding to $f=2$ and $2.5$  \citep{Brandenburg_Kahniashvili_2017, Bhat_Zhou_Loureiro_2021, Dwivedi2024}. In addition, the helical component of the magnetic spectrum exhibits a slope of $k^3$. These inertial range slopes develop naturally despite the different nature in the fall of the spectrum in our initial condition. Interestingly, $k^{-2}$ in the nonhelical turbulence corresponds to weak turbulence \citep{Tobias_Cattaneo_Boldyrev_2012}. The rising part of the initial magnetic energy spectrum is chosen to scale as $k^4$, \textit{i.e.} $r=4$, motivated by primordial magnetic fields generated during early-universe phase transitions \citep{Christensson_Hindmarsh_Brandenburg_2001,Durrer_Caprini_2003}. Previous studies have explored the effect of initializing the magnetic spectrum with values of $r$ different from $4$ and found that inverse transfer of magnetic energy persists in these cases as well albeit with perhaps less efficiency \citep{Brandenburg_Sharma_Vachaspati_2023}. 

\begin{figure}
    \centering
    \includegraphics{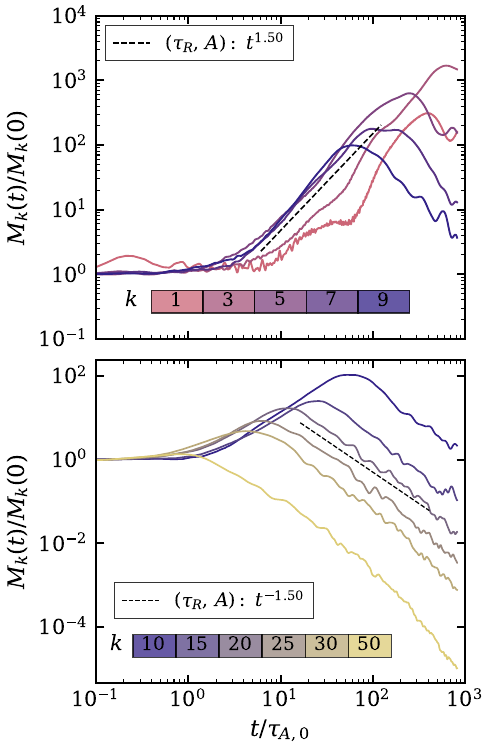}
    \caption{Temporal evolution of the rising (\textbf{top}) and decaying (\textbf{bottom}) spectral modes in a strict 2D $2048^2$ simulation with initial $S \simeq 1000$ and $k_p = 60$. The measured scalings are consistent with the reconnection timescale and with anastrophy conservation.}
    \label{fig:2D S1000 spec evol}
\end{figure}
We use high resolution simulations to measure the power law indices associated with the temporal evolution of the spectral modes and compare with the theoretical expectation as given in \Eq{eqn: defn of sf sr}. 
These exponents can be used to discern the relevant combination of decay timescale quantified by $n$ and the conserved quantity represented by $\alpha$. 

As an example, we mention the specific case of nonhelical turbulence. For anastrophy conservation, $\alpha=1$ and we note that irrespective of $n$, we obtain the following, 
\begin{equation}
    M(k,t) \sim \begin{cases}
        k^4 t^{3/2}, & k<k_p(t)\\
        k^{-2} t^{-3/2}, & k\geq k_p(t)
    \end{cases}.
    \label{ana-exp}
\end{equation}
Note that $n$ appears in the formulae specified in \Eq{eqn: defn of sf sr} in a product along with $(\alpha -1)$ and thus, $n$ doesn't contribute. 
In the strict 2D case, where we have only two components of the field, there can be simply no other candidate for a conserved quantity, in $\eta \to 0$ limit, other than anastrophy. Consequently, we expect the scaling predicted in \Eq{ana-exp} to hold robustly. This is indeed the case, as can be seen in \Fig{fig:2D S1000 spec evol}. 
We find that the mode evolution extracted from both rising and falling parts of the spectra exhibits an excellent match with the theoretical expectation of $\sigma_r=\sigma_f=1.5$. This provides strong support for the validity of our broken power-law model.
\begin{figure}
    \centering
    \includegraphics{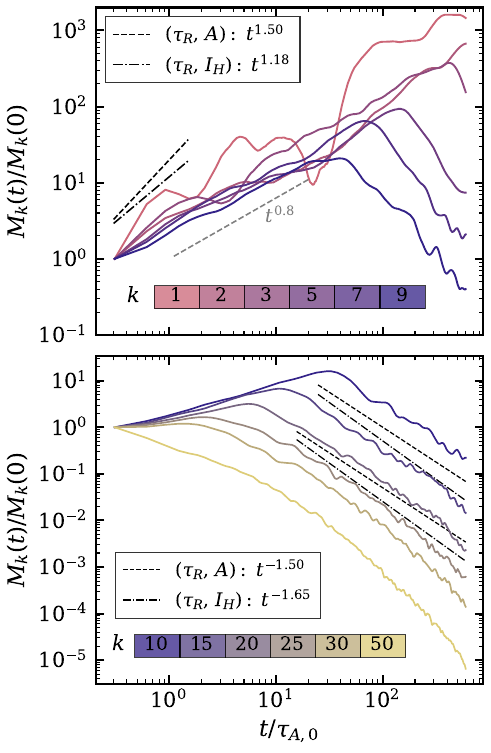}
    \caption{Same as Fig.~\ref{fig:2D S1000 spec evol}, but for the 2.5D nonhelical case.}
    \label{fig:2.5D nonhel spec evol}
\end{figure}

\begin{figure}
    \centering
    \includegraphics{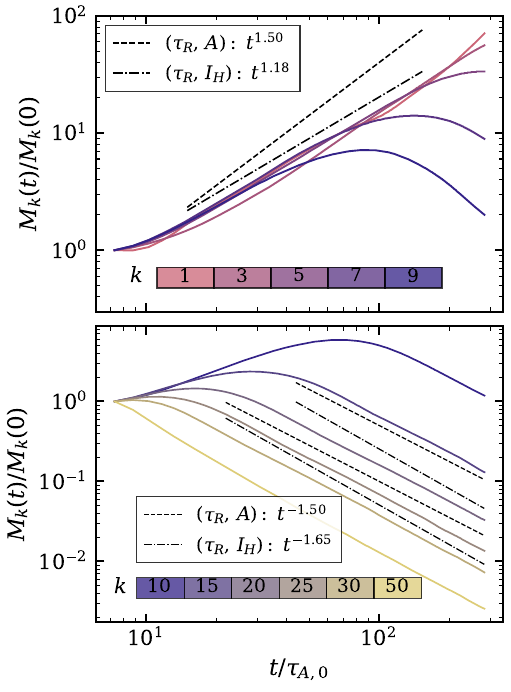}
    \caption{Same as Fig.~\ref{fig:2D S1000 spec evol}, but for the 3D nonhelical case with initial $S\approx2300$. Run from \cite{Brandenburg_Kahniashvili_2017}}
    \label{fig:3D nonhel spec evol}
\end{figure}

Since our derivation leading to \Eq{eqn:time evol of energy} makes no assumptions about the dimensionality, the same scaling laws are expected to apply in 2.5D (all three components of the field but is allowed to evolve in a 2D domain) and 3D. However, with  helicity fluctuations, $I_H$ integral conservation ($\alpha=4/5$) leads to $\sigma_r=1.18$ and $\sigma_f=1.65$.  

To probe the spectral evolution scalings efficiently, we require sufficiently large initial $k_p$ for two reasons: (i) to obtain $t^{-1}$ scaling for the magnetic energy for longer time period (before either box scale effects or resistive effects take care) and (ii) to obtain relevant modes that transition from sub-inertial to inertial range. Accordingly, we have chosen initial $k_p=60$ in most cases.
We show the spectral evolution of the individual modes in 2.5D and 3D nonhelical runs in \Figs{fig:2.5D nonhel spec evol}{fig:3D nonhel spec evol}. We find that, in 2.5D, the very small $k$ modes from the rising spectral range are a bit erratic in their temporal behaviour, whereas modes at $k=3$ and above seem to show altogether a very different $\sigma_r \sim 0.8$. However, in the rising parts of the 3D case, the early evolution is steep, closer to $\{\tau_R,I_A\}$ scaling, whereas the latter evolution seems closer to $\{\tau_R,I_H\}$.

On the other hand, in the 2.5D case, we find the evolution of modes from the falling spectral range agrees very well with theoretical expectation of anastrophy conservation. In particular, the evolution of modes $k=10, 15, 20$ and $25$ follow scaling from anastrophy conservation, while $k=30$ and $50$ seem to be deviating from a straight line substantially. 
Similar agreement is observed in the 3D case as well in the temporal evolution of falling or inertial range modes. Fig~\ref{fig:3D nonhel spec evol} illustrates this behaviour for modes till $k=30$, whereas, the $k=50$ mode deviates away from a straight line evolution. 

Next, we turn to the helical case. Since magnetic helicity is not well defined in strictly two-dimensional systems, we restrict our analysis to the 2.5D and 3D configurations.

\begin{figure}
    \centering
    \includegraphics{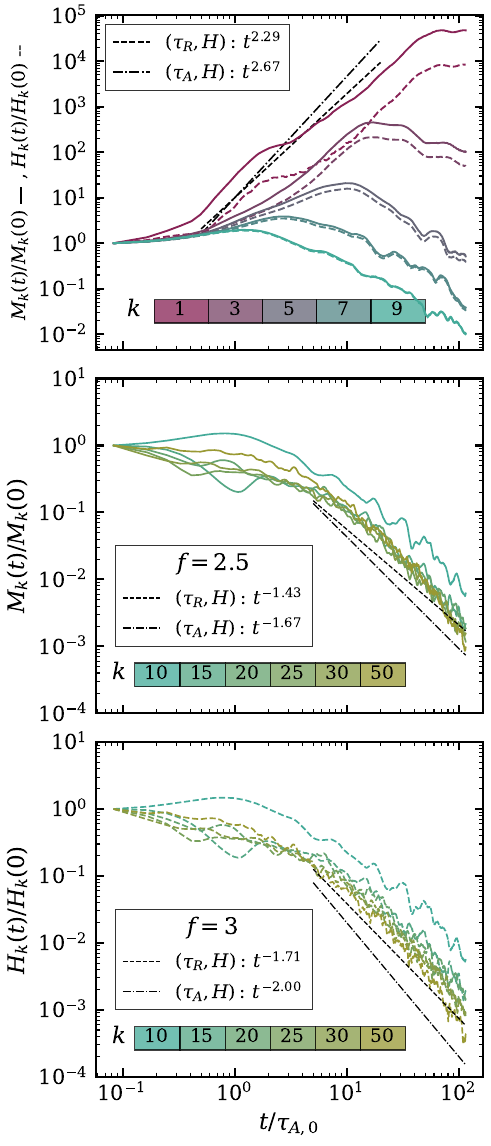}
    \caption{
Same as Fig.~\ref{fig:2D S1000 spec evol}, but for the 2.5D helical case.
The measured scalings are consistent with the reconnection timescale
and helicity conservation.
}
    \label{fig:2.5D hel spec evol}
\end{figure}
Previous studies have shown that in the 2.5D case, even when the system is initialized with fully helical magnetic fields, a nonhelical component is generated during the evolution (see Fig. 3 of \cite{Dwivedi2024}). This behaviour is not observed in fully three-dimensional simulations. Consequently, for a meaningful comparison with theoretical predictions, we focus specifically on the helical component of the total magnetic energy, particularly in the 2.5D case and 
it has been confirmed that this helical part of the magnetic energy decays as $E_M^{\rm hel}(t) \sim t^{-4/7}$ \citep{Dwivedi2024}. This behaviour is consistent with a {$n=0.5, \alpha=2$} model corresponding to the decay timescale being governed by SP reconnection, and magnetic helicity being the relevant conserved quantity. However, there has been some ambiguity in the literature regarding the precise value of the decay exponent. In particular, an alternative scaling of $p = 2/3$ corresponds to $n = 0$, i.e. a decay timescale set by the Alfvénic time \citep{Brandenburg_Kahniashvili_2017}. Next, we present the spectral evolution of helical magnetic turbulence, which provides additional and independent evidence to distinguish between these competing models with different $n$.

In \Fig{fig:2.5D hel spec evol}, we present the results from the 2.5D helical simulation. We show both the total magnetic energy spectrum, $M(k)$, and the spectrum associated with the helical component, $M_H(k)=k~H(k)/2$. Note that while both these spectra: $M(k)$ and  $M_H(k)$ have the same slope in the rising sub-inertial part of the spectrum i.e. $r=4$, their slopes differ in the falling part of the spectrum. In particular, we observe $f=2.5$ in the case of $M(k)$ and  $f=3$ in the case of $M_H(k)$ (see Appendix~\ref{sec:MkHkspectra}). 

The top panel of \Fig{fig:2.5D hel spec evol} shows the temporal evolution of the low-$k$ modes. We find that the growth of power at those wavenumbers is fairly steep, and therefore more consistent with the reconnection-driven decay model, which predicts a steeper power-law scaling than the Alfvénic model. The Alfvénic prediction shows partial agreement at early times, most clearly for the $k=2$ mode. This behaviour is expected, as the early-time evolution is dominated by transient dynamics governed by the Alfvén timescale as shown by the weak scaling of $C$ with $S$ in a previous section~\ref{sec:timescale}. 

The middle and bottom panels of \Fig{fig:2.5D hel spec evol} correspond to large $k$ modes (in the inertial range) from $M(k)$ and $M_H(k)$, respectively. 
We find that while the match with  reconnection model is remarkably good in the 3D case while it is reasonable in the 2.5D case, providing strong support for the reconnection-driven model. 

\begin{figure}
    \centering
    \includegraphics{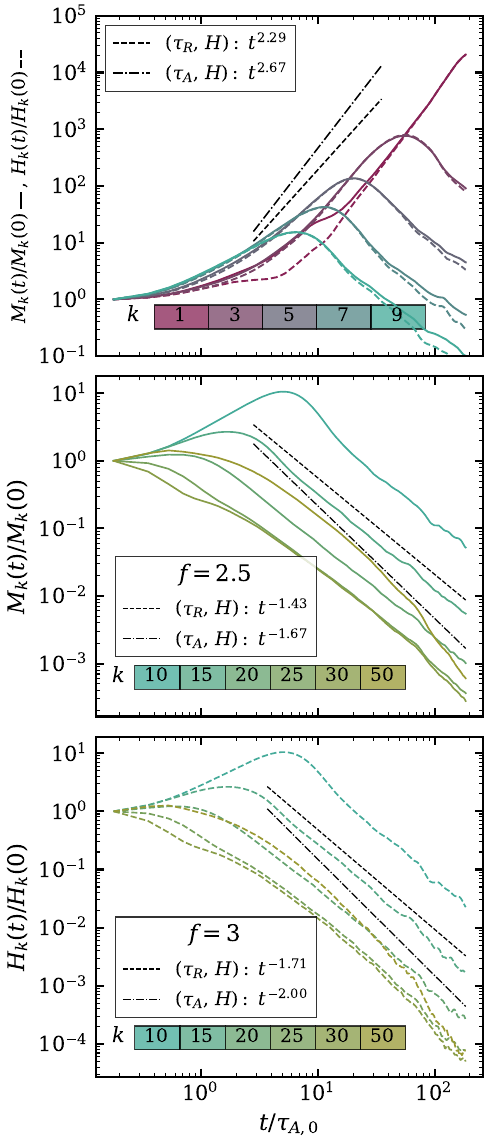}
    \caption{
Same as Fig.~\ref{fig:2.5D hel spec evol}, but for the 3D nonhelical
$1024^3$ simulation with initial $S \approx 1000$ and $k_p = 25$.
}

    \label{fig:3D hel spec evol}
\end{figure}

Finally, we present the results from the 3D case in \Fig{fig:3D hel spec evol}. Note that the nonhelical component of total energy is subdominant, and therefore the evolution from both total energy $M(k)$ and the helical part $M_H(k)$ is expected to agree robustly with the theoretical expectations. While the agreement at low wavenumbers, shown in the top panel, is not very clear, the reconnection-based model provides a better match to the data in both the middle and bottom panels.

Overall, we find that the evolution of sub-inertial modes is not very reliable diagnostic due to noisy evolution but the inertial range modes agree very well with the theoretical predictions. 


\section{Resolving reconnection layers in a turbulent system\label{sec:recres}}

    The results of previous sections, along with previous works \citep{Bhat_Zhou_Loureiro_2021, Hosking_Schekochihin_2021, Dwivedi2024}, indicate persuasively that magnetic reconnection is at play in magnetically dominated decaying turbulence. Reconnection dynamics are controlled by current sheets in the system. In particular, the flux-frozen condition breaks down with the reconnecting current-sheet layer. Thus, it is important to resolve these current sheets across their thickness. 

Since the magnetic spectra in these decaying systems are peaked, the dominant structures at any given time are expected to be at the peak scale.
A simplistic expectation is that the current sheets corresponding to these interacting structures would be of similar size. We incorporate this consideration while estimating the smallest scale in problem that must be resolved in order to capture reconnection dynamics correctly.

In the Sweet–Parker reconnection regime relevant to our simulations, the current-sheet aspect ratio is controlled by $S$.
Let the current sheet length be given by $L$.
At any given time, the magnetic energy is peaked at scales $k_p(t)\sim 1/\xi_M(t)$. We can naively assume $L\sim \xi_M$. According to SP model, the current sheet width is given by,
\begin{equation}
    \label{sp cs width}
    \delta = S^{-1/2}L
\end{equation}
Let the smallest length scale that can be resolved in a simulation with $N^{d}$ grid points (d: dimensions) and system size $L_S$ be $\Delta x$. 
The current-sheet width must then be at least $1.5\,\Delta x$ to be properly resolved. 
Defining $f_{\mathrm{CS}} \equiv \delta / (1.5\,\Delta x)$, the criterion for ensuring that current sheets are resolved is

\begin{equation}
    \label{csres}
    \delta \geq 1.5\, \Delta x \text{ \quad \textit{i.e.}\quad } f_{CS} \ge 1
\end{equation} 

\begin{table}[h!]
\centering
\begin{tabular}{|c|c|c|}
\hline
\textbf{Paper} & \textbf{Sim ID} & $\delta / (1.5\, \Delta x)$ \\
\hline
\parbox{4cm}{\centering\citet{Zhou_Sharma_Brandenburg_2022}} & K60D1bc & 0.11 \\
\parbox{4cm}{\centering\citet{Hosking_Schekochihin_2021}}  & R23 & 0.30 \\
\parbox{4cm}{\centering\citet{Armua_2023}} & $NH_{pz0}$ & 3.02 (low S) \\
\parbox{4cm}{\centering\citet{Bhat_Zhou_Loureiro_2021}} & $A3D$ & 0.86 \\
\hline
\end{tabular}

\caption{Simulation parameters and diagnostics for different runs.}
\label{tab:sim_summary}
\end{table}

\begin{figure}[h!]
    \centering
    \includegraphics{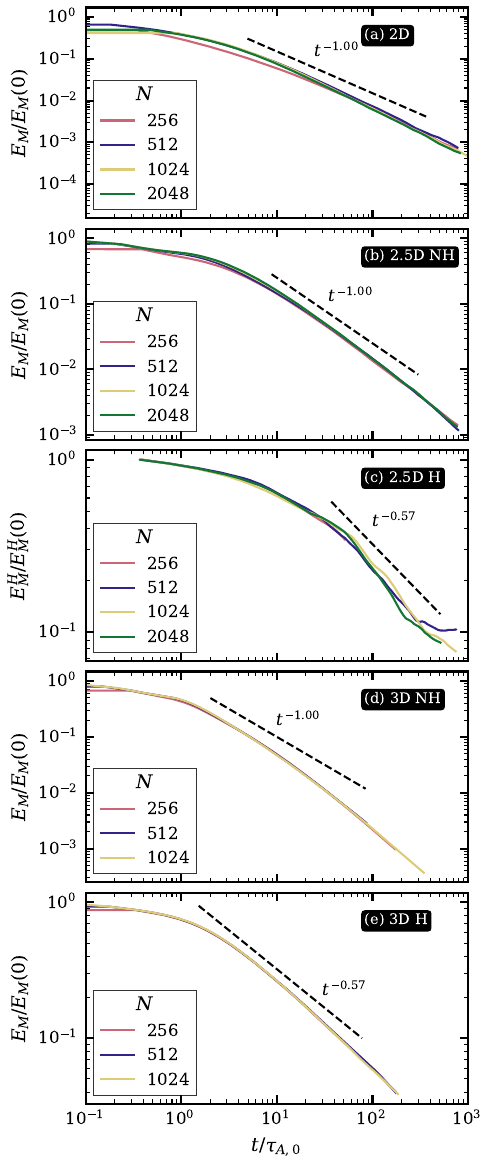}
    \caption{Temporal evolution of magnetic energy. Each panel corresponds to a different configuration, as indicated by the black boxes. Curves within each panel show simulations at varying numerical resolutions, demonstrating resolution-independent evolution. 
    }
    \label{fig:EMvst}
\end{figure}

\begin{figure*}
    \centering
    \includegraphics{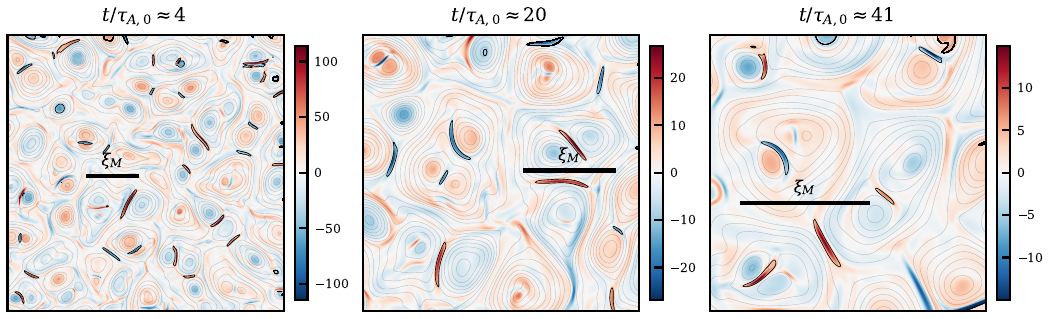}
    \caption{
Evolution and size of current sheets in strict 2D simulations.
Panels display the current density $J_z$ with overlaid contours
of the magnetic vector potential $A$. Current sheets included in the Minkowski Functionals  analysis are marked by black contours, while thick black lines indicate the correlation scale at each time.
}     
\label{fig:cs length}
\end{figure*}

In the literature, the temporal power-law decay exponent of the magnetic field in several earlier simulations (including some of the runs listed in Table~\ref{tab:sim_summary}) was found to agree with reconnection-driven theoretical expectations, irrespective of whether the criterion in \Eq{csres} was satisfied. 
Therefore, the validity of this criterion is called into question. 
To assess the criterion, we perform a suite of simulations (as listed in Table~\ref{tab:sim_params}) at different numerical resolutions while keeping the microphysical resistivity fixed, thereby effectively varying $f_{CS}$.

In \Fig{fig:EMvst}, we present the resulting magnetic energy evolution for runs with identical initial Lundquist number $S$ (i.e. the same resistivity $\eta$, as well as identical initial Alfvén speed $V_A$ and peak wavenumber $k_p$), but differing spatial resolution. The magnetic energy curves from these simulations lie nearly on top of one another. The five panels, from top to bottom, correspond to the five distinct categories defined by dimensionality and helicity.

Within each category, we observe that reducing the resolution and consequently reducing the current-sheet resolution factor $f_{\rm CS}$, does not lead to any significant change in the energy decay behaviour. In particular, the inferred decay exponent $p$ remains unchanged across resolutions. This result is striking, as one would naively expect that if reconnection physics governs the dynamics, inadequate resolution of current sheets should strongly affect the decay.

One possible explanation is that the effective Lundquist numbers associated with individual current sheets are substantially smaller than the global Lundquist number $S$ defined using system integrated quantities. If reconnection proceeds in locally low-$S$ current sheets, then the apparent insensitivity of the decay to $f_{\rm CS}$ would naturally follow.

 The natural next step is to examine local current-sheet structures in real space. Our aim is to characterize their geometric properties: specifically, their sizes relative to the magnetic correlation scale, across simulations with varying numerical resolutions, with particular emphasis on the lower-resolution runs where the global resolution criterion appears to fail.
\begin{figure*}
    \centering
    \includegraphics[width=0.32\linewidth]{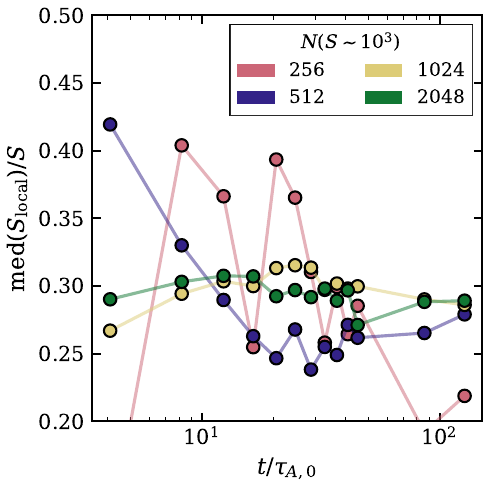}
    \includegraphics[width=0.32\linewidth]{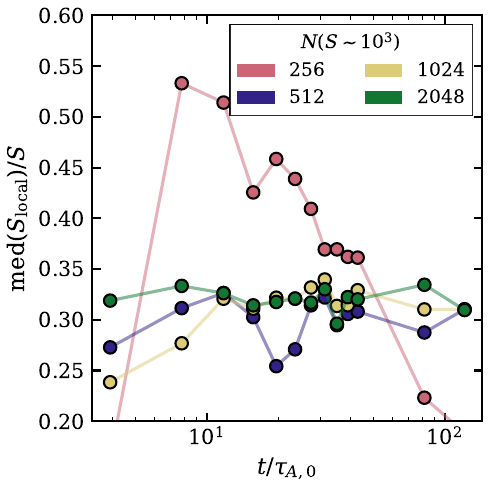}
        \includegraphics[width=0.32\linewidth]{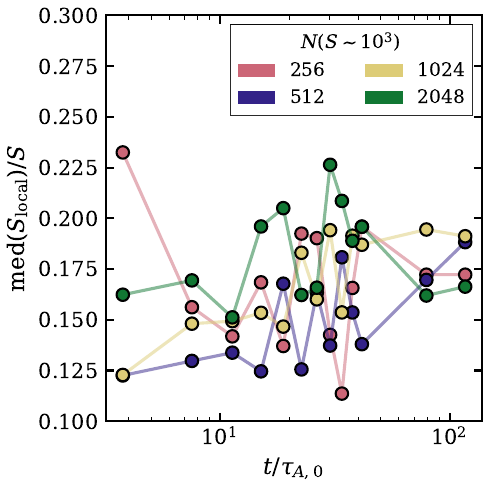}
  \caption{Time evolution of the ratio of the median current-sheet Lundquist number to
the global Lundquist number for the 2D, 2.5D nonhelical, and 2.5D helical cases,
showing that local Lundquist numbers are smaller than the global value.} 
    \label{fig:s-global-local-ratio}
\end{figure*}

Such an analysis requires a quantitative and robust method for measuring the dimensions of current sheets. One suitable tool, previously employed by us in \cite{Dwivedi2024}, is the use of Minkowski functionals (first applied in cosmology by \citet{Mecke_Buchert_Wagner_1994}; see references therein), which provide a systematic way to characterize the morphology of spatial structures. Since reliable identification of current sheets is most straightforward in two dimensions, we restrict this analysis to two-dimensional data. We have previously applied the Minkowski analysis to 3D data in \citet{Dwivedi2024} to quantify the quasi–two-dimensional nature of 3D MHD turbulence.

\begin{table*}
\centering
\begin{tabular}{lccccccccccccc}
\hline
$N$ & $t=1$ & $t=2$ & $t=3$ & $t=4$ & $t=5$ & $t=6$ & $t=7$ & $t=8$ & $t=9$ & $t=10$ & $t=11$ & $t=21$ & $t=31$ \\\hline

\multicolumn{14}{c}{\textbf{2D Nonhelical}}\\\hline
\hline
256 & \cellcolor[HTML]{f98e52} - & \cellcolor[HTML]{f98e52} -0.049 & \cellcolor[HTML]{cbe982} -0.256 & \cellcolor[HTML]{fff7b2} -0.194 & \cellcolor[HTML]{f98e52} 0.060 & \cellcolor[HTML]{f98e52} 0.097 & \cellcolor[HTML]{f98e52} 0.062 & \cellcolor[HTML]{f98e52} 0.055 & \cellcolor[HTML]{f98e52} 0.011 & \cellcolor[HTML]{f98e52} -0.014 & \cellcolor[HTML]{f98e52} -0.020 & \cellcolor[HTML]{fff7b2} -0.196 & \cellcolor[HTML]{a5d86a} -0.315 \\
512 & \cellcolor[HTML]{f98e52} 0.110 & \cellcolor[HTML]{f98e52} 0.049 & \cellcolor[HTML]{f98e52} -0.020 & \cellcolor[HTML]{fee08b} -0.143 & \cellcolor[HTML]{ecf7a6} -0.224 & \cellcolor[HTML]{75c465} -0.365 & \cellcolor[HTML]{a5d86a} -0.307 & \cellcolor[HTML]{a5d86a} -0.345 & \cellcolor[HTML]{75c465} -0.386 & \cellcolor[HTML]{75c465} -0.397 & \cellcolor[HTML]{3faa59} -0.447 & \cellcolor[HTML]{75c465} -0.383 & \cellcolor[HTML]{3faa59} -0.410 \\
1024 & \cellcolor[HTML]{ecf7a6} -0.206 & \cellcolor[HTML]{75c465} -0.359 & \cellcolor[HTML]{75c465} -0.362 & \cellcolor[HTML]{75c465} -0.358 & \cellcolor[HTML]{a5d86a} -0.334 & \cellcolor[HTML]{3faa59} -0.403 & \cellcolor[HTML]{75c465} -0.391 & \cellcolor[HTML]{3faa59} -0.434 & \cellcolor[HTML]{75c465} -0.389 & \cellcolor[HTML]{75c465} -0.383 & \cellcolor[HTML]{3faa59} -0.407 & \cellcolor[HTML]{3faa59} -0.430 & \cellcolor[HTML]{75c465} -0.398 \\
2048 & \cellcolor[HTML]{75c465} -0.362 & \cellcolor[HTML]{a5d86a} -0.347 & \cellcolor[HTML]{3faa59} -0.401 & \cellcolor[HTML]{75c465} -0.399 & \cellcolor[HTML]{75c465} -0.388 & \cellcolor[HTML]{3faa59} -0.444 & \cellcolor[HTML]{138c4a} -0.460 & \cellcolor[HTML]{138c4a} -0.455 & \cellcolor[HTML]{138c4a} -0.451 & \cellcolor[HTML]{138c4a} -0.476 & \cellcolor[HTML]{138c4a} -0.455 & \cellcolor[HTML]{3faa59} -0.405 & \cellcolor[HTML]{138c4a} -0.474 \\

\hline

\multicolumn{14}{c}{\textbf{2.5D Nonhelical}}\\\hline
\hline
256 & \cellcolor[HTML]{f98e52} - & \cellcolor[HTML]{ecf7a6} -0.226 & \cellcolor[HTML]{cbe982} -0.290 & \cellcolor[HTML]{ecf7a6} -0.220 & \cellcolor[HTML]{f98e52} 0.054 & \cellcolor[HTML]{f98e52} 0.066 & \cellcolor[HTML]{f98e52} 0.043 & \cellcolor[HTML]{f98e52} 0.049 & \cellcolor[HTML]{f98e52} 0.027 & \cellcolor[HTML]{f98e52} -0.002 & \cellcolor[HTML]{f98e52} 0.037 & \cellcolor[HTML]{fee08b} -0.138 & \cellcolor[HTML]{cbe982} -0.278 \\
512 & \cellcolor[HTML]{f98e52} 0.009 & \cellcolor[HTML]{f98e52} 0.002 & \cellcolor[HTML]{fdbb6c} -0.055 & \cellcolor[HTML]{f98e52} -0.049 & \cellcolor[HTML]{fee08b} -0.143 & \cellcolor[HTML]{ecf7a6} -0.248 & \cellcolor[HTML]{cbe982} -0.298 & \cellcolor[HTML]{a5d86a} -0.323 & \cellcolor[HTML]{a5d86a} -0.326 & \cellcolor[HTML]{a5d86a} -0.340 & \cellcolor[HTML]{a5d86a} -0.336 & \cellcolor[HTML]{75c465} -0.380 & \cellcolor[HTML]{75c465} -0.362 \\
1024 & \cellcolor[HTML]{75c465} -0.357 & \cellcolor[HTML]{75c465} -0.370 & \cellcolor[HTML]{75c465} -0.373 & \cellcolor[HTML]{a5d86a} -0.314 & \cellcolor[HTML]{a5d86a} -0.327 & \cellcolor[HTML]{75c465} -0.355 & \cellcolor[HTML]{cbe982} -0.297 & \cellcolor[HTML]{cbe982} -0.293 & \cellcolor[HTML]{a5d86a} -0.337 & \cellcolor[HTML]{a5d86a} -0.350 & \cellcolor[HTML]{75c465} -0.384 & \cellcolor[HTML]{3faa59} -0.404 & \cellcolor[HTML]{3faa59} -0.404 \\
2048 & \cellcolor[HTML]{75c465} -0.355 & \cellcolor[HTML]{75c465} -0.374 & \cellcolor[HTML]{3faa59} -0.406 & \cellcolor[HTML]{3faa59} -0.403 & \cellcolor[HTML]{3faa59} -0.409 & \cellcolor[HTML]{3faa59} -0.407 & \cellcolor[HTML]{3faa59} -0.412 & \cellcolor[HTML]{3faa59} -0.431 & \cellcolor[HTML]{a5d86a} -0.317 & \cellcolor[HTML]{75c465} -0.382 & \cellcolor[HTML]{138c4a} -0.453 & \cellcolor[HTML]{138c4a} -0.476 & \cellcolor[HTML]{138c4a} -0.509 \\
\hline

\multicolumn{14}{c}{\textbf{2.5D Helical}}\\\hline
\hline
256 & \cellcolor[HTML]{fee08b} -0.139 & \cellcolor[HTML]{fdbb6c} -0.056 & \cellcolor[HTML]{fdbb6c} -0.070 & \cellcolor[HTML]{a5d86a} -0.326 & \cellcolor[HTML]{fff7b2} -0.151 & \cellcolor[HTML]{a5d86a} -0.325 & \cellcolor[HTML]{3faa59} -0.415 & \cellcolor[HTML]{a5d86a} -0.337 & \cellcolor[HTML]{f98e52} 0.006 & \cellcolor[HTML]{75c465} -0.381 & \cellcolor[HTML]{138c4a} -0.535 & \cellcolor[HTML]{138c4a} -0.520 & \cellcolor[HTML]{138c4a} -0.520 \\
512 & \cellcolor[HTML]{75c465} -0.362 & \cellcolor[HTML]{a5d86a} -0.335 & \cellcolor[HTML]{3faa59} -0.422 & \cellcolor[HTML]{75c465} -0.400 & \cellcolor[HTML]{3faa59} -0.406 & \cellcolor[HTML]{3faa59} -0.413 & \cellcolor[HTML]{75c465} -0.385 & \cellcolor[HTML]{75c465} -0.384 & \cellcolor[HTML]{75c465} -0.391 & \cellcolor[HTML]{a5d86a} -0.344 & \cellcolor[HTML]{3faa59} -0.408 & \cellcolor[HTML]{138c4a} -0.466 & \cellcolor[HTML]{a5d86a} -0.304 \\
1024 & \cellcolor[HTML]{75c465} -0.365 & \cellcolor[HTML]{a5d86a} -0.340 & \cellcolor[HTML]{75c465} -0.367 & \cellcolor[HTML]{75c465} -0.350 & \cellcolor[HTML]{75c465} -0.390 & \cellcolor[HTML]{3faa59} -0.439 & \cellcolor[HTML]{75c465} -0.378 & \cellcolor[HTML]{75c465} -0.384 & \cellcolor[HTML]{138c4a} -0.457 & \cellcolor[HTML]{75c465} -0.363 & \cellcolor[HTML]{3faa59} -0.408 & \cellcolor[HTML]{75c465} -0.375 & \cellcolor[HTML]{138c4a} -0.469 \\
2048 & \cellcolor[HTML]{3faa59} -0.422 & \cellcolor[HTML]{138c4a} -0.473 & \cellcolor[HTML]{138c4a} -0.454 & \cellcolor[HTML]{138c4a} -0.467 & \cellcolor[HTML]{138c4a} -0.470 & \cellcolor[HTML]{138c4a} -0.469 & \cellcolor[HTML]{138c4a} -0.513 & \cellcolor[HTML]{138c4a} -0.519 & \cellcolor[HTML]{138c4a} -0.509 & \cellcolor[HTML]{138c4a} -0.523 & \cellcolor[HTML]{138c4a} -0.535 & \cellcolor[HTML]{3faa59} -0.414 & \cellcolor[HTML]{138c4a} -0.601 \\
\hline
\vspace{0.1 cm}
\end{tabular}

\begin{tikzpicture}
  \def\n{11}      
  \def\w{1.0}     
  \def\h{0.25}    
  \def\ytick{-0.12} 

  \definecolor{c0}{HTML}{f98e52}
  \definecolor{c1}{HTML}{fdbb6c}
  \definecolor{c2}{HTML}{fee08b}
  \definecolor{c3}{HTML}{fff7b2}
  \definecolor{c4}{HTML}{ecf7a6}
  \definecolor{c5}{HTML}{cbe982}
  \definecolor{c6}{HTML}{a5d86a}
  \definecolor{c7}{HTML}{75c465}
  \definecolor{c8}{HTML}{3faa59}
  \definecolor{c9}{HTML}{138c4a}
  \definecolor{c10}{HTML}{006837}

  \foreach \i in {0,...,10} {
    \pgfmathsetmacro\x{\i*\w}
    \fill[color=c\i] (\x,0) rectangle ++(\w,\h);
  }

  \draw (0,0) rectangle (\n*\w,\h);

  \foreach \i/\lab in {
    0/{$>$0.00}, 1/{-0.05}, 2/{-0.10}, 3/{-0.15}, 4/{-0.20}, 
    5/{-0.25}, 6/{-0.30}, 7/{-0.35}, 8/{-0.40}, 9/{-0.45}, 10/{-0.50}, 11/{$<$-0.55}
  }{
    \pgfmathsetmacro\x{\i*\w}
    \draw (\x,0) -- ++(0,-0.06);   
    \node[below] at (\x,\ytick) {\scriptsize \lab};
  }
\end{tikzpicture}


\caption{
Fit values of $\log(\delta/L)$ versus $\log S$ for various simulations and
times. Well-resolved current sheets—those at late times and/or higher
resolutions—exhibit a clear SP scaling.
}
\label{cs-slopes}
\end{table*}
The procedure for computing Minkowski functionals, along with validation and benchmarking tests, is described in Appendix~\ref{bench}. Details of the current-sheet identification algorithm itself are provided in Appendix~\ref{csid}.

With these tools in place, we now proceed to analyse the strict 2D and 2.5D simulations.
We begin by visualizing the current-sheet structures in these systems and comparing their characteristic lengths with the magnetic integral scale, $\xi_M$, at the same instant in time. The purpose of this comparison is to assess whether the typical extent of current sheets is indeed set by the large-scale magnetic structure, as is often implicitly assumed.
In \Fig{fig:cs length}, we show contours of the current density, outlined in black. The current sheets appear as thin, elongated filaments in the plane; however, given that the third dimension is a symmetry direction, these structures are in fact sheet-like. Superposed on these plots as thick black lines are segments representing the magnetic integral scale, $\xi_M$, at the corresponding times.
A clear difference is evident: the overlaid $\xi_M$ markers are significantly longer than the observed current-sheet lengths at the same time. This disparity persists across the snapshots shown, indicating that the characteristic length of current sheets is substantially smaller than the magnetic integral scale.
We therefore conclude that the simple, or naive, expectation that current-sheet lengths are comparable to the integral scale, or to the size of the dominant magnetic structures, is not borne out by the simulations.

From the analysis above, we can now conclude that estimates of the Lundquist number based on global quantities—specifically $\sqrt{2E_M}$, which is spatially integrated, and the magnetic integral scale $\xi_M$—are not representative of the current sheets that actually govern the dynamics of the system. The reconnection processes that control the decay are instead mediated by localized structures whose characteristic scales differ substantially from the global magnetic scale.
Motivated by this, we directly measure the lengths of individual current sheets in the domain and combine these with the local magnetic field strength to define a local Lundquist number $S_{\rm local}=B_{\rm local}l/\eta$. In \Fig{fig:s-global-local-ratio}, we show the temporal evolution of the ratio of the average (or median) local Lundquist number, $S_{\rm local}$, to the global Lundquist number, $S$, for the 2D and 2.5D simulations. In all cases, this ratio is systematically different from unity, indicating a persistent disparity between local and global measures of the Lundquist number.

The left and middle panels of \Fig{fig:s-global-local-ratio} show that, in the nonhelical cases, $S_{\rm local}/S$ settles to a value of approximately $\sim 0.3$. Moreover, the agreement improves with increasing numerical resolution, suggesting that this ratio is converged. The right panel corresponds to the helical case, for which the ratio is somewhat smaller, in the range $\sim 0.175$–$0.2$.

We find that this reduction in the local Lundquist number relative to the global value arises almost entirely from the difference in the characteristic length scale. While the global definition of $S$ employs $L = \xi_M$, the local current-sheet length $L_{\rm local}$ is smaller by a factor of approximately $3$–$4$, as already can be gleaned from \Fig{fig:cs length}. 

\begin{figure}
    \centering
    \includegraphics[width=0.9\linewidth]{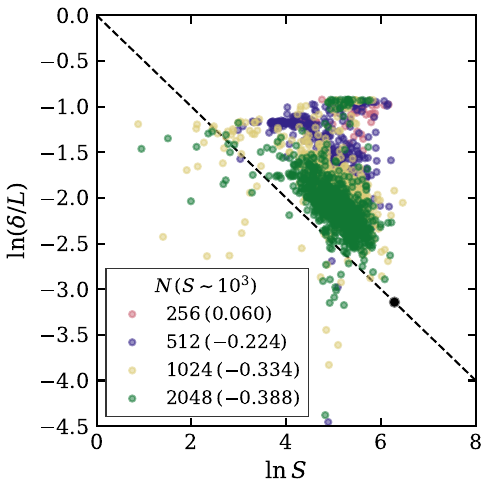}
        \includegraphics[width=0.9\linewidth]{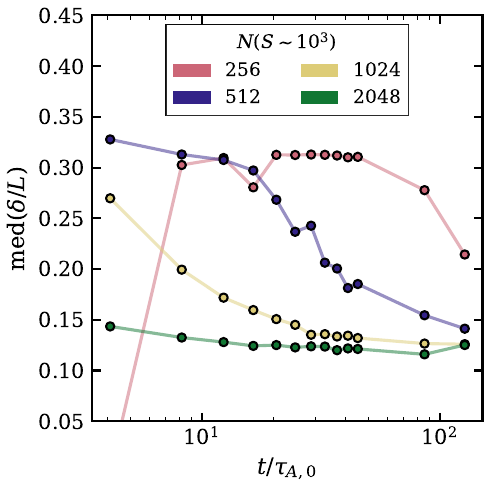}
    \caption{Plot of $\ln\, \left(\delta / L\right)$ with $\ln\, S$ at $t=5$ to check for SP scaling. 
    Lower panel - Medians of current sheet lengths and widths as a function of resolution for various times showing a `sense' of convergence at large resolutions.}
    \label{fig:2D SP scaling convergence}
\end{figure}
Next, we present the results of our analysis of local CSs properties as a function of time. The widths and lengths of current sheets are estimated using Minkowski functionals (MFs), following the procedure described in Appendix~\ref{csid}.
Using these measurements, we examine the scaling of current-sheet aspect ratios by fitting $\ln(\delta/L)$ as a function of $\ln(S)$. The resulting slope directly yields the exponent $-n$. The fitted slopes for all three cases—strict 2D nonhelical, 2.5D nonhelical, and 3D helical runs—are shown in Table~\ref{cs-slopes}.

At very early times, when the system is still dominated by transient dynamics, reliable measurement of current-sheet dimensions is difficult due to the small size and rapid evolution of structures. As the system evolves and more coherent structures grow, the current-sheet statistics become increasingly robust. Correspondingly, the extracted slope values, shown in Table~\ref{cs-slopes}, begin to converge toward the Sweet–Parker reconnection prediction of $n = 0.5$, with convergence improving systematically at higher numerical resolutions.
This scaling is not recovered at the lowest resolution ($256^2$), and is only weakly evident in the $512^2$ simulations. Clear agreement with the Sweet–Parker scaling emerges only once the current sheets are sufficiently well resolved.

In \Fig{fig:2D SP scaling convergence}, we show the data points from local CS analysis in the strict 2D run, at $t=5$, graphically with $\ln{(\delta/L)}$ on ordinate axis and $\ln{S_{\rm local}}$ on the abscissa and how it compares visually to the $-n=-0.5$ slope. 
In the top panel, we indicate in the legend the extracted slope (in bracket) from local CS analysis from runs with different resolutions. As can be noted very distinctively the trend is that with increasing resolution $n$ increases. We believe its possible that the value would get closer to $0.5$ with better and better resolutions. 

\begin{figure}
    \centering
    \includegraphics[width=0.9\linewidth]{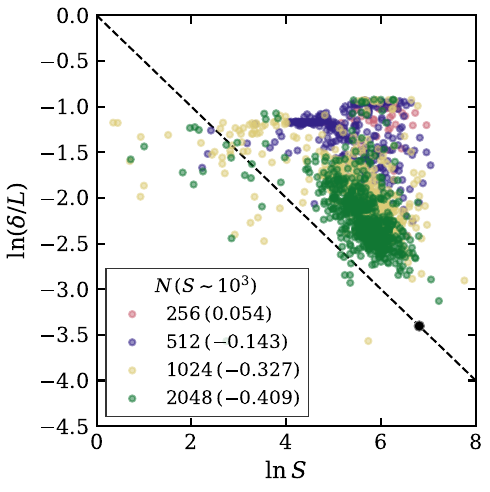}
    \includegraphics[width=0.9\linewidth]{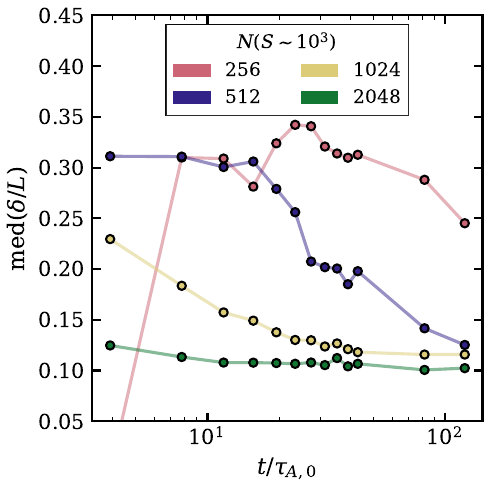}
    
    \caption{Same as Fig \ref{fig:2D SP scaling convergence} but for 2.5D nonhelical case.}
    \label{fig:2.5D nonhel CS-scal}
\end{figure}
\begin{figure}
    \centering
    \includegraphics[width=0.9\linewidth]{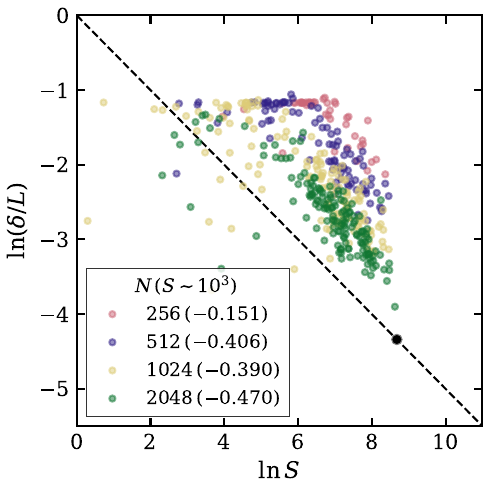}

    \includegraphics[width=0.9\linewidth]{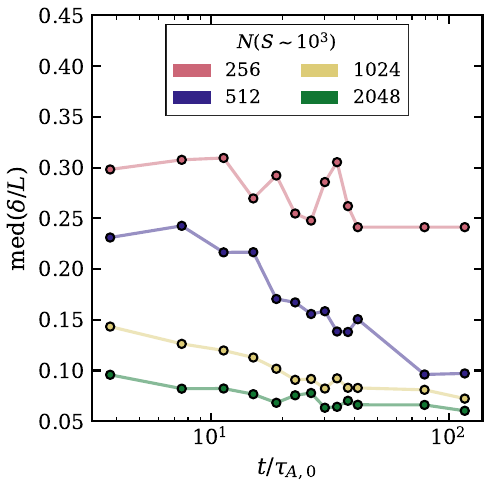}
    
    \caption{Same as Fig \ref{fig:2D SP scaling convergence} but for 2.5D helical case.}
    \label{fig:2.5D hel CS-scal}
\end{figure}
While \Fig{fig:EMvst} demonstrated that the global magnetic energy decay curves do not exhibit an obvious sensitivity to reduced current-sheet resolution (smaller $f_{\rm CS}$), we find that scaling measurements of individual current sheets require substantially higher resolution to converge. In particular, the current-sheet aspect ratio shown in the lower panel of \Fig{fig:2D SP scaling convergence} exhibits clear convergence only in the two highest-resolution runs, $1024^2$ and $2048^2$.

Also the value of aspect ratio  is consistent with our finding that $S/S_{\rm local} \sim 3$. Using the Sweet–Parker (SP) reconnection model, the expected aspect ratio of a local current sheet is
 $\delta/L=S_{\rm local}^{-1/2}$.
 With $S_{\rm local} \simeq 0.3
~S$, this yields $\delta/L\sim 0.1$, in good agreement with the values observed in \Fig{fig:2D SP scaling convergence}.
Thus, although global decay properties appear somewhat insensitive to resolution, accurate characterization of local current-sheet geometry, and hence direct verification of reconnection scalings, demands sufficiently high numerical resolution.

In \Figs{fig:2.5D nonhel CS-scal}{fig:2.5D hel CS-scal}, we show similar results for the cases of 2.5D nonhelical run and helical run respectively. 
We see that the median slope values are around $-0.41$ and $-0.47$ in the highest resolution runs again highlighting the importance of resolution for a reconnection-driven system. And as before, the current sheet aspect ratios seem to converge better in the two higher resolution runs with $1024^2$ and $2048^2$. 

A current-sheet analysis for the 3D case, similar to that performed above for the 2D and 2.5D cases, was carried out in our previous work \citep{Dwivedi2024}.
In particular, the evolution of current-sheet lengths for 3D helical and non-helical cases shown in Figs.~24 and 25 of that work demonstrates that the scaling $\delta/L \sim S_{\mathrm{local}}^{-0.5}$ is satisfied to within a factor of two.
This indicates that, even in the 3D case, local current sheets are broadly consistent with Sweet--Parker scaling.

\section{Discussions and Conclusions}

\subsection{Decay timescale}
In the literature on decaying MHD turbulence, magnetic reconnection has only relatively recently been recognized as the possible primary mechanism governing the decay, robustly supplanting the earlier view that the decay timescale is set by the Alfvénic time \citep{Bhat_Zhou_Loureiro_2021}. This reconnection-based picture also naturally accounts for the presence of inverse transfer in both magnetically helical and nonhelical systems \citep{Dwivedi2024}. Physically, the coalescence of magnetic structures via reconnection progressively leads to the formation of larger-scale magnetic structures. At low Lundquist numbers, the relevant regime for such decaying turbulence has been identified as Sweet–Parker reconnection.

However, in \cite{Brandenburg_Neronov_Vazza_2024} it was reported that the decay timescale scaling, $\tau_{\rm decay} \sim S^n \tau_A$, did not follow the Sweet–Parker expectation of $n = 1/2$, but instead yielded a shallower scaling closer to $n \simeq 1/3$. This result was obtained in a two-dimensional setting, where reconnection is expected to be particularly robust and therefore theoretically clean.

In the present work, we revisited this analysis using a combination of global decay diagnostics and direct measurements of local current-sheet properties. We find that the inferred value of $n$ depends sensitively on the method used to characterize the relevant length scale and calculate the Lundquist numner. When the analysis relies on global measures such as the magnetic integral scale, the extracted scaling can be biased. By contrast, approaches based on either fitted energy evolution along with the conserved quantity constraint or direct local current-sheet diagnostics consistently recover the Sweet–Parker scaling $n = 1/2$ in both 2D/2.5D and 3D simulations.
It is possible that the discrepancy reported in earlier work arises primarily from the use of the integral scale in the calculations. Owing to the discrete evolution of the correlation scale in numerical simulations, the integral scale does not reliably track the characteristic size of reconnecting structures. 

\subsection{Conserved quantity in decaying MHD turbulence}
While it is well established that helical MHD turbulence is constrained by the conservation of magnetic helicity, the situation in the nonhelical case is being debated. The question is whether the relevant constraining conserved quantity is anastrophy or  helicity fluctuations quantifying integral $I_H$. A striking feature of nonhelical MHD turbulence is that inverse transfer of energy emerges robustly only in simulations with sufficiently high numerical resolution. This sensitivity to resolution explains why inverse transfer in nonhelical systems was identified only relatively recently, around 2015 \citep{Brandenburg_Kahniashvili_Tevzadze_2015}, and why its discovery was initially unexpected.

The reconnection-based interpretation of inverse transfer was first motivated by observations in 2D/2.5D systems, where inverse transfer arises naturally through the coalescence of magnetic islands \citep{Zhou_Bhat_Loureiro_Uzdensky_2019}. In these settings, the relevant conserved quantity is clearly anastrophy, which constrains the dynamics. Note that this is fundamentally different from the hydrodynamic case, where the nature of inverse transfer depends strongly on dimensionality. In two-dimensional hydrodynamics, inverse cascades arise because kinetic energy is more robustly conserved than enstrophy, whereas in three dimensions there is no inverse transfer of energy. Instead, small-scale structures decay first, leading to just an apparent growth of coherence length scales \citep{Davidson_2015,Panickacheril_John_Donzis_Sreenivasan_2022}.

MHD differs crucially in this respect: inverse transfer is observed in both 2D and 3D. A natural implication is that the physical mechanism responsible in 2D is also operative in 3D. This interpretation is further supported by the fact that inverse transfer in 3D becomes apparent only at high resolutions, where the flow exhibits partial quasi-two-dimensionalization, allowing anastrophy-like constraints to influence the dynamics \citep{Dwivedi2024}.
Consistent with this picture, we find that the temporal power-law scalings in 2D and 3D agree closely over an intermediate time interval, after the initial transient phase but before resistive and box-scale effects dominate the 3D evolution. Nevertheless, because the interpretation of inverse-transfer slopes in fully 3D systems remains somewhat ambiguous, an alternative viewpoint has been proposed in which the dynamics are constrained instead by $I_H$.

It is, first, highly unlikely that the physical agent responsible for inverse energy transfer in three dimensions is entirely distinct from that operating in two dimensions. More fundamentally, when multiple quantities are approximately conserved, it is essential to distinguish between those that actively constrain the dynamics and those that are merely conserved without exerting strong dynamical control.
The helicity-fluctuation integral, $I_H$, is intended to quantify the net helicity contained in localized patches of the flow. However, even in systems with non-zero mean helicity, a meaningful helicity-based constraint requires the magnetic field to be close to fully helical—that is, the vector potential and magnetic field must be fully aligned. This condition is necessary for helicity to act as a tight dynamical constraint. No analogous requirement has been established for the applicability of $I_H$, making its role as a controlling invariant less clear.
Moreover, it has been explicitly shown that the anastrophy integral decays more slowly than $I_H$, implying that anastrophy remains dynamically relevant over longer timescales. Consistent with this expectation, our analysis of spectral mode evolution in the inertial range shows significantly better agreement with predictions based on anastrophy than with those derived from $I_H$.

\subsection{Local current sheet analysis}
We have found that the evolution of the magnetic energy is largely independent of the numerical resolution of the simulations. At first sight, this appears puzzling, since magnetic reconnection has been  identified as the likely physical mechanism driving the decay, and therefore the expectation is that current sheets associated with the peak-energy scale magnetic structures must be adequately resolved. Our local current sheet analysis resolves this apparent contradiction.

We find that the magnetic correlation scale is not the correct measure of the length of the current sheets. Instead, the current sheets are systematically smaller than the correlation scale by, approximately, a factor of  $3-4$. As a consequence, the effective Lundquist number relevant for reconnection is reduced to about one-third to one-fourth of the global Lundquist number estimated using the correlation scale. This reduction explains why the decay of magnetic energy exhibits little sensitivity to numerical resolutions.

A local analysis of current sheet properties further 
recovers the exponent $n=0.5$ for current sheet aspect ratio scaling with Lundquist number, particularly in simulations with higher resolutions and at later times, where possible under-resolution effects are minimized. When combined with the global decay analysis, this provides substantial and self-consistent evidence that the decay is indeed reconnection-driven. Our reassurance in this conclusion stems from the fact that these two diagnostics are fundamentally different in nature: one probes the global temporal evolution of integrated quantities, while the other measures local properties of current sheets at fixed times.

A series of studies, including \citet{Zhdankin_Uzdensky_Perez_Boldyrev_2013} and \citet{Servidio_Matthaeus_Shay_Cassak_Dmitruk_2009}, have examined local current sheet statistics in the context of forced MHD turbulence, where the absence of Sweet–Parker scaling was attributed to the strong distortion of current sheets by turbulent flows. In contrast, the present work focuses on freely decaying turbulence, where such distortions are significantly less severe, allowing the reconnection physics to manifest more clearly.

An additional outcome of our analysis is the identification of the magnetic correlation scale with the size of merging magnetic islands, as seen in Fig~\ref{fig:cs length}. For an island of characteristic size $2R$, the correlation scale is found to be approximately $4R$, while the current sheet length is of order $R$, bounded above by the island diameter. This geometric interpretation naturally explains the factor of $3-4$ separation between the correlation scale and the current sheet length observed in the simulations. A more controlled investigation of this connection using isolated magnetic island–merging simulations is left for future work. Overall, this demonstrates that local current sheet properties can differ substantially from global measures such as the correlation scale, while still governing the global decay of magnetic energy. 

\subsection{Implications for early universe}
\begin{figure}[h!]
    \centering
    \includegraphics{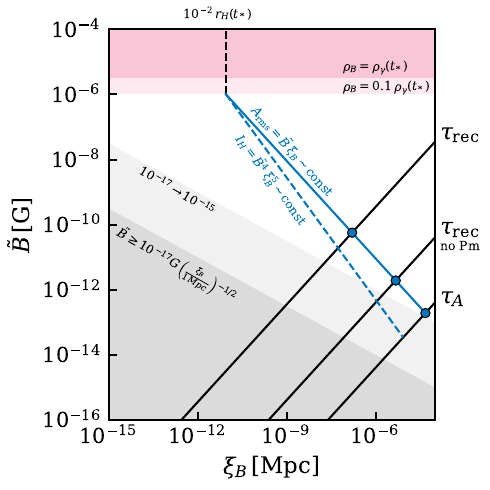}
    \caption{Evolution of the comoving magnetic field strength \(\tilde{B}\) as a function of the correlation scale \(\xi_B\) in the early Universe. The gray shaded region indicates parameter space excluded by observational constraints, while the pink shaded region denotes the region excluded by initial conditions. The pink regions are calculated assuming that a fraction of the total energy of the Universe at the electroweak phase transition (\(\rho\)) went into the magnetic field (\(\rho_B\), here taken to be 1/10). The vertical dashed line indicates the correlation length of the initial magnetic field, taken to be \(1/100\) of the Hubble radius at the electroweak phase transition. The black curves show the implicit relations between \(\tilde{B}\) and \(\xi_B\) at recombination, obtained by equating different decay timescales to the cosmological time, as indicated along the right-hand axis. The blue curve corresponds to evolution constrained by anastrophy conservation, with its intersections with the recombination curves marked by blue circles.}
    \label{fig:BlB_plot}
\end{figure}

Our results, together with previous studies of decay timescales, have direct implications for the evolution of primordial magnetic fields.  
We have observational bounds mainly from $\gamma$-ray studies in cosmic voids \citep{Neronov_Vovk_2010}.
These constraints are commonly visualized using a comoving magnetic field-correlation scale(\(\tilde{B}\!-\!\xi_B\)) diagram, as shown in Fig.~\ref{fig:BlB_plot}. In this representation, decay timescales proportional to \(\xi_B/\tilde{B}\), when equated to the cosmological time, define implicit relations between the magnetic field strength \(\tilde{B}\) and the correlation scale \(\xi_B\) evaluated at recombination, and therefore appear as increasing curves in the \(\tilde{B}\!-\!\xi_B\) plane. Conservation laws, by contrast, impose explicit decreasing relations, such as \(\tilde{B}\,\xi_B \simeq \mathrm{const}\). The intersection of these increasing and decreasing curves determines the magnetic field strength and coherence scale at recombination.

Recently it was shown that reconnection-driven decay 
can satisfy current observational bounds on intergalactic magnetic fields, whereas Alfvénic decay leads only to a marginal clearance \citep{Hosking_Schekochihin_2023}. Given the significant uncertainties associated with these observations, it is preferable that any clearance of observational bounds be robust rather than marginal.

In the reconnection-driven scenario considered by \citet{Hosking_Schekochihin_2023}, the decay timescale is increased by its dependence on the magnetic Prandtl number $Pm$,
\begin{equation}
    \label{eq:reconnection with Pm}
    \frac{\tau_{\rm rec}}{\tau_A}
    \sim
    \left(1+Pm\right)^{1/2}\,
    \text{min}\left(S^{1/2},\,S_c^{1/2}\right).
\end{equation}

This enhancement implies a slower decay of the magnetic field and allows a larger field strength \(\tilde{B}\) to survive until recombination. In the early Universe, where the Spitzer magnetic Prandtl number is extremely large,
\(
Pm^{\rm Sp} \sim 10^7 \left(\frac{T}{0.3~\rm eV}\right),
\)
and the Lundquist number satisfies \(S \gg S_c\), this effect leads to a robust clearance of observational bounds.

However, the validity of this scaling in the asymptotic regime relevant to the early Universe—characterized by simultaneously large magnetic Prandtl number and vanishing diffusivities—is not guaranteed. The \(Pm\) dependence entering the decay timescale is a natural extrapolation of results obtained for \(Pm \sim 1\), but it may break down when both viscosity and resistivity become asymptotically small. In particular, the width of a Sweet--Parker current sheet is predicted to scale as
\(
\delta \propto Pm^{1/4}
\) \citep{Comisso_Grasso_2016}.
If one imagines holding \(Pm\) fixed while reducing both diffusivities such that the diffusive length scales become much smaller than the current-sheet width, the dynamics within the sheet should no longer depend on the diffusivities. This contradicts a persistent \(Pm\)-dependent current-sheet structure and casts doubt on whether the Prandtl-number dependence of the reconnection rate can survive in the cosmological limit.

The recent work of \citet{Brandenburg_Neronov_Vazza_2024} provides support for the above argument. In particular, using the results of their numerical simulations (see Section~7 of their paper), they find that the decay timescale does not exhibit a systematic dependence on the magnetic Prandtl number, even at large \(Pm\) and small diffusivities. This lends additional weight to the possibility that the Prandtl-number dependence inferred from moderate-\(Pm\) and diffusivity studies may not persist in the asymptotic regimes relevant for the early Universe.

Motivated by this concern, we also reconstruct the \(\tilde{B}\!-\!\xi_B\) diagram assuming anastrophy conservation and compare with approximate \(I_H\) conservation in Fig~\ref{fig:BlB_plot}. We find that the  decay on a reconnection timescale without magnetic Prandtl number dependence leads only to almost the same level of clearance of observational constraints as that by $I_H$ line when $Pm$ dependence is included. Thus, the slower decay due to anastrophy conservation clears observational bounds with a more conservative estimate of $\tau_{rec}$, which $I_H$ constrained decay does not. 
These results indicate that conclusions regarding the survival of magnetic fields generated during the electroweak phase transition (EWPT)—used here as a representative example—should be interpreted with caution, as they depend sensitively on the assumed decay mechanism and its asymptotic validity.

\section*{Data availability}
The scripts supporting the findings of this study are openly available at \url{https://github.com/chandranathan/SpecEvol}. The datasets used in the current study are available from the corresponding author on reasonable request.
\begin{acknowledgments}
CA thanks the members of the AstroPlasma@ICTS group for their valuable help and discussions. PB would like to thank Axel Brandenburg and Andrii Neronov for helpful discussions. PB thanks JinLin Han for the invitation to the International Symposium on Cosmic Magnetic Field held in Beijing, China. We acknowledge project RTI4001 of the Dept. of Atomic Energy, Govt. of India. The simulations were performed on the ICTS HPC cluster Contra.
\end{acknowledgments}

%

\software{PENCIL CODE \citep{Pencilcode}}

\appendix
\section{Magnetic Energy and Helicity Spectra \label{sec:MkHkspectra}}

\begin{figure}[h!]
    \centering
    \includegraphics{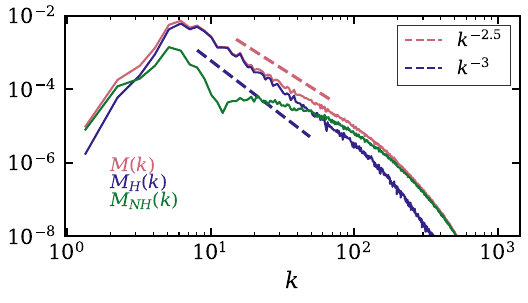}
    \includegraphics{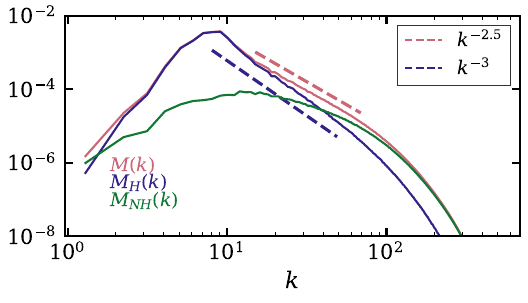}
    \caption{Magnetic energy spectra $M(k)$, helical $M_H(k)$, and non-helical energy components shown as functions of wavenumber $k$ at $t=5$ for helical 2.5D (\textbf{left}) and 3D (\textbf{right}) simulations. Dashed lines indicate inertial-range reference power-law scalings, $M(k)\sim k^{-2.5}$ and $M_H(k)\sim k^{-3}$.
}
    \label{fig:hel HkMk evolution}
\end{figure}
\section{Evolution of integral length scale \label{sec:evol of lint}}

\begin{figure}[h!]
    \centering


    \begin{subfigure}{0.32\textwidth}
        \centering
        \includegraphics{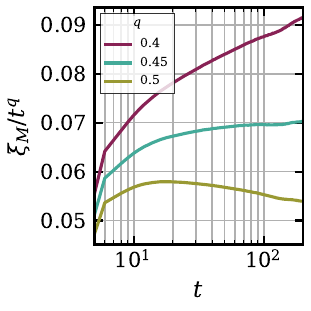}
    \end{subfigure}
    \hfill
    \begin{subfigure}{0.32\textwidth}
        \centering
        \includegraphics{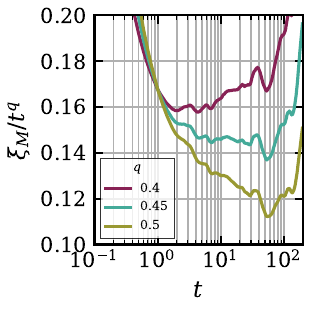}
    \end{subfigure}
    \hfill
        \begin{subfigure}{0.32\textwidth}
        \centering
        \includegraphics{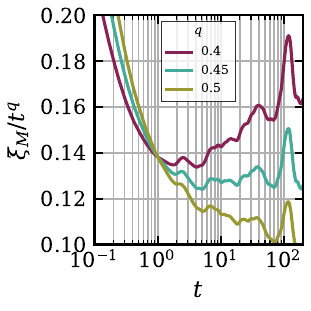}
    \end{subfigure}
    
    \vspace{0.5em}

      \begin{subfigure}{0.32\textwidth}
        \centering
        \includegraphics{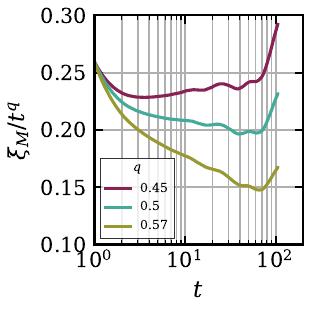}
    \end{subfigure}
    \hfill
    \begin{subfigure}{0.32\textwidth}
        \centering
        \includegraphics{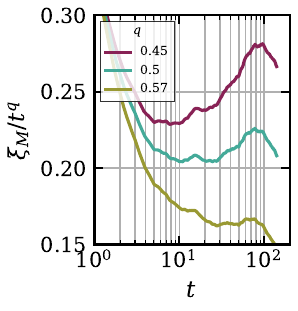}
    \end{subfigure}
    \hfill
    \begin{subfigure}{0.32\textwidth}
    \centering
    \phantom{\rule{\linewidth}{0.6\linewidth}}

\end{subfigure}
    \hfill

    \caption{
Compensated evolution of $L_{\mathrm{int}}$, showing that it does not capture the correct length-scale scaling.
\textbf{Top}: non-helical runs (expected $q = 0.5$); \textbf{bottom}: helical runs (expected $q = 0.57$).
\textbf{Left to right}: 3D, 2.5D, and strictly 2D.
    }
    \label{fig:evolution of lint}
\end{figure}

The evolution of the integral length scale of the magnetic energy is shown in \Fig{fig:evolution of lint}.
For the helical cases, the integral length scale is calculated using the helical component of the magnetic field.
It is clear from the figure that the expected length-scale scalings are not reproduced by $L_{\mathrm{int}}$.

\section{Benchmarking 2D Minkowski Functionals}
\label{bench}

To estimate the length and width of current sheets in two-dimensional turbulent simulations, we use Minkowski functionals. In two dimensions, there are three Minkowski functionals: area, perimeter, and connectivity. To benchmark our algorithm, which uses the \textsc{skimage.measure} module from the scikit-image package \citep{vanderWalt2014}, we construct ellipses with varying ratios of the minor ($a$) to major ($b$) axes. We then apply the algorithm to compute the area and perimeter, treating each ellipse as a single connected component. From these quantities, we subsequently obtain the Minkowski functionals, characteristic length scales, and derived geometric measures.

\begin{figure}[h!]
    \centering
    \includegraphics{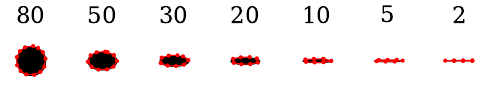}
    \caption{Contours of ellipses identified using \textsc{skimage.measure.find\_contours}, with a fixed major axis length ($=80$) and varying minor axis lengths.}
    \label{fig:benchmark contour}
\end{figure}

\begin{figure}[h!]
    \centering
    \includegraphics[scale=0.95]{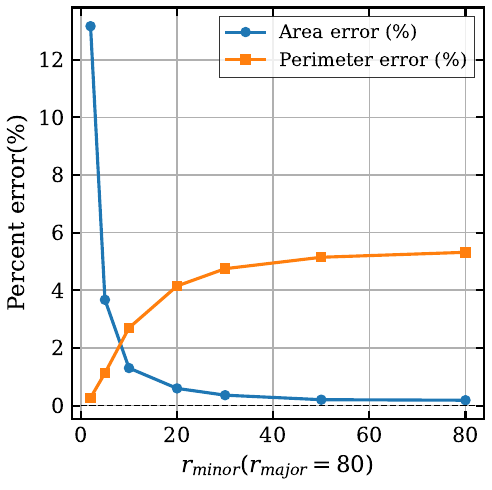}
    \caption{Minkowski functional diagnostics showing errors in the calculated area and perimeter.}
    \label{fig:benchmark error and F}
\end{figure}

The algorithm correctly captures the contours of the ellipses, as shown in Fig.~\ref{fig:benchmark contour}, and returns an ordered list of points along each contour. The area ($V_0$) is computed using the Shoelace formula, while the perimeter ($V_1$) is obtained by summing the distances between successive contour points. Assuming connected surfaces, the two characteristic length scales derived from these Minkowski functionals are
\begin{equation}
    \label{eq:2D_MF}
    \tilde{l}_1 = \frac{V_0}{\pi V_1}, \quad \text{and} \quad \tilde{l}_2 = \frac{V_1}{2\pi}.
\end{equation}
To assess the accuracy of the algorithm, we compare the computed area and perimeter with their theoretical values for an ellipse. The area and perimeter are given by
\begin{equation}
    \label{eq:area perimeter of ellipse}
    A = \pi ab, \quad 
    P \approx \pi\left[3(a+b) - \sqrt{(3a+b)(a+3b)}\right],
\end{equation}
where the expression for the perimeter uses Ramanujan’s approximation. The errors in both the area and perimeter estimates decrease with increasing length scale, particularly for larger widths. Although the perimeter exhibits a small systematic asymptotic offset, all errors remain below five percent (Fig.~\ref{fig:benchmark error and F}).

The length scales defined in Eq.~\ref{eq:2D_MF} do not directly correspond to the semi-minor and semi-major axes of the ellipse. To understand their geometric meaning, we approximate Eq.~\ref{eq:2D_MF} in the limit of large aspect ratio ($b/a \gg 1$). This limit is physically relevant, as current sheets are expected to be highly elongated structures. Retaining only the leading-order terms yields $\tilde{l}_1 \simeq 1.58\,a$ and $\tilde{l}_2 \simeq 0.63\,b$. Accordingly, the semi-minor and semi-major axes can be estimated as
\begin{equation}
    \label{eq:semi-minor axis}
    l_1 = \frac{2\,\tilde{l}_1}{1.58},
\end{equation}
\begin{equation}
    \label{eq:semi-major axis}
    l_2 = \frac{2\,\tilde{l}_2}{0.63}.
\end{equation}

\begin{figure}[h!]
    \centering
    \includegraphics{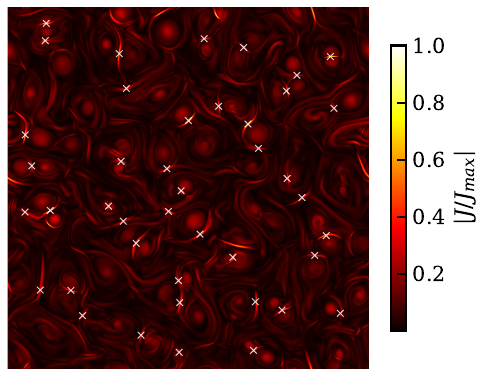}
    \includegraphics{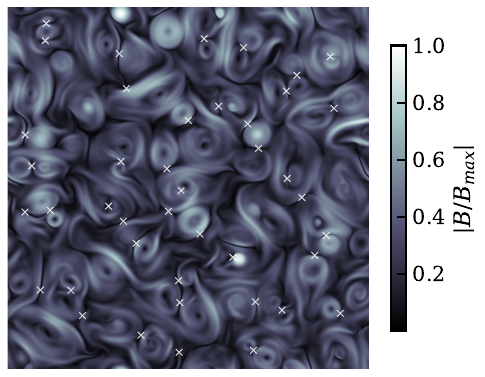}
    \caption{Identification of current sheets in a 2.5D nonhelical $2048^2$ simulation. Plots of current density (\textit{left}) and magnetic field (\textit{right}); white crosses mark the locations of the identified current sheets.}
    \label{fig:current sheet identification}
\end{figure}

\section{Current Sheet identification in turbulent simulations}
\label{csid}

In the previous section, we have verified that our algorithm correctly estimates Minkowski Functionals for 2D density fields. Now, we need to pass the ``correct" current density fields to quantify reconnecting current sheets. To identify these current sheets, we employ the same philosophy as we did for the 3D case in \citet{Dwivedi2024}. We apply a global threshold (taken to be $4$) using global $J_{rms}$ to find regions of large current density. Then, using a lower threshold, we identify current sheets and consider all their centroids. In a square of size $\xi_M/4$, we consider the current sheet with the maximum $J$ value as the representative current sheet.   A sample implementation for a 2.5D simulation is shown in \ref{fig:current sheet identification}.

For each representative current sheet, we consider the current density field centered at its centroid, with size $\xi_M/2$. The Minkowski Functionals are calculated for the representative current sheet. The length $L$ and width $\delta$ of a current sheet, approximated as a highly elongated ellipse (see Fig.~\ref{fig:cs length}), are identified with the semi-major axis (\Eq{eq:semi-major axis}) and the semi-minor axis (\Eq{eq:semi-minor axis}), respectively.
 Using the local $B_{rms}$, which is close to the global $B_{rms}$, we calculate the local Lundquist number as $S = B_{rms, local}\, L/\eta$.

\section{Varying the current density threshold}
\begin{figure}
    \centering
    \includegraphics[width=0.4\linewidth]{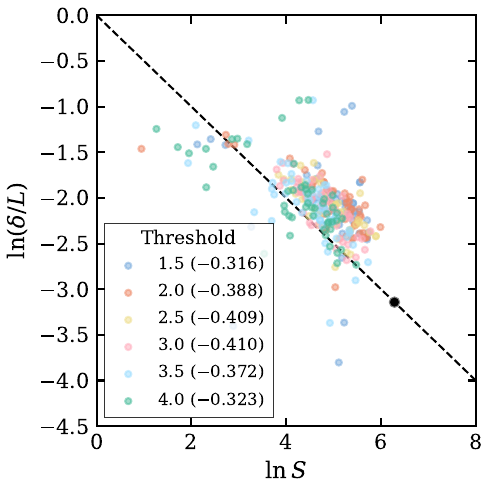}
    \caption{Plot of $log\, \left(\delta / L\right)$ with $log\, S$ to check for Sweet-Parker scaling at various thresholds.}
    \label{thresh}
\end{figure}
In \Fig{thresh}, the exercise of extracting the slope is done with different thresholds on the current density for CS identification. For example, a threshold value of 3 corresponds to current sheets that are above three times $J_{\rm rms}$ for marking the relevant contour for MF analysis. We find that increasing the threshold increases the extracted value of $n$ from 0.31 to 0.4, but beyond a certain value of threshold, $n$ ends up decreasing, likely because the number of pixels decreases, and that decreases the fidelity of the MF calculation. 

\section{Local $B_{\rm rms}$ and $L$ of the current sheets}
\begin{figure*}[h!]
  \centering

  \includegraphics[width=0.3\textwidth]{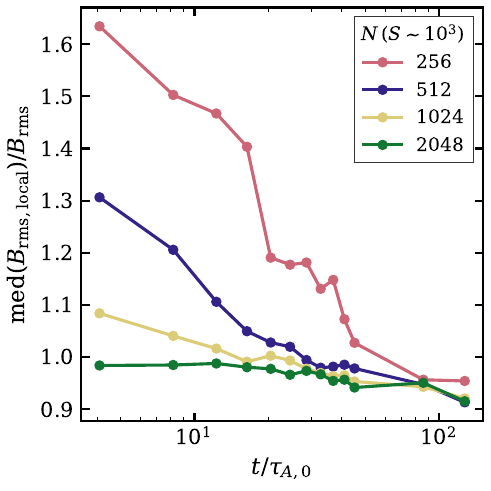}\hfill
  \includegraphics[width=0.3\textwidth]{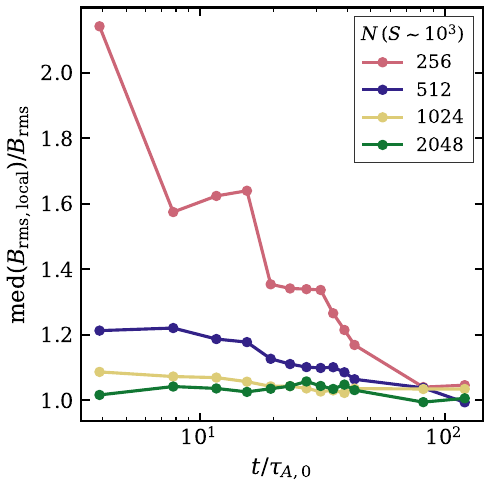}\hfill
  \includegraphics[width=0.3\textwidth]{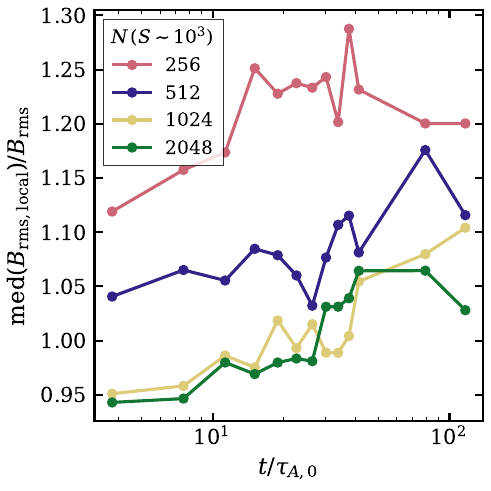}

  \medskip

    \includegraphics[width=0.3\textwidth]{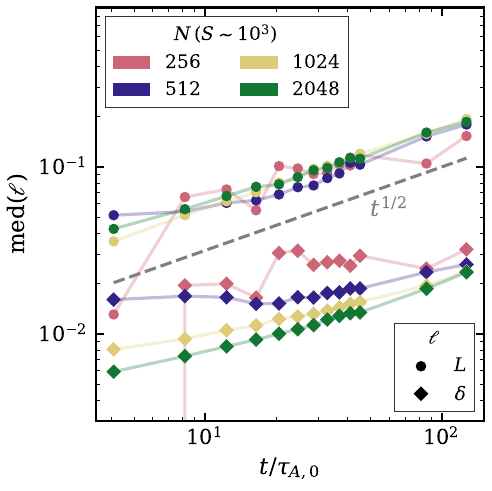}\hfill
  \includegraphics[width=0.3\textwidth]{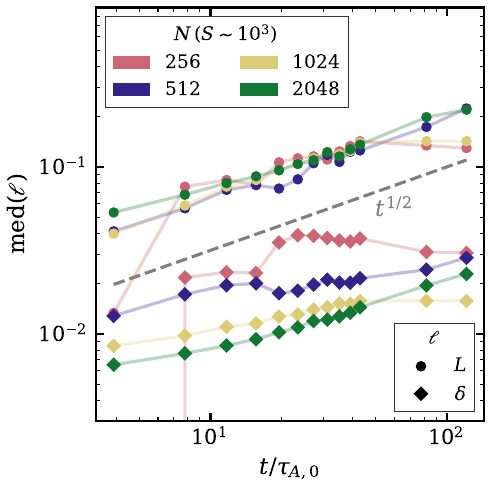}\hfill
  \includegraphics[width=0.3\textwidth]{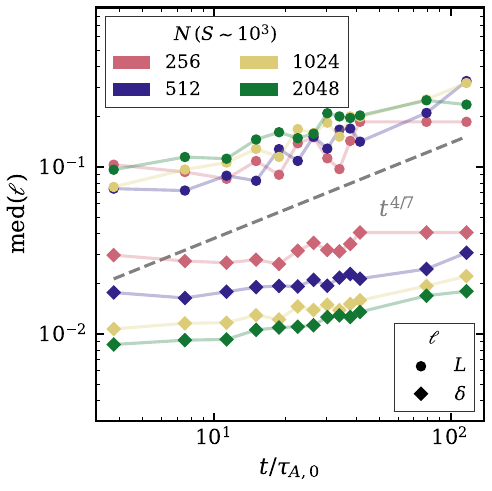}
    
    \caption{From left to right, the columns correspond to the 2D, 2.5D nonhelical,
and 2.5D helical cases, respectively.
\textbf{Top row:} Time evolution of the ratio of the median 
$B_{\rm rms}$ in the vicinity of the sheets to the global $B_{\rm rms}$.
\textbf{Bottom row:} Time evolution of the median current-sheet length $L$
and width $\delta$, indicating scalings consistent with the reconnection
timescale and anastropy (or helicity, for the helical case) conservation.
} 
  \label{fig:local B L}
\end{figure*}
We showed in Section~\ref{sec:recres} that the local Lundquist numbers are smaller than
the global Lundquist number. Figure~\ref{fig:local B L} shows that the local
$B_{\rm rms}$ in the vicinity of current sheets is comparable to the global
$B_{\rm rms}$ for all cases (top panel). In the main text, we presented the
ratios of the current-sheet length to width; here, we instead show the full
time evolution of the current-sheet dimensions. In the nonhelical case, the
current-sheet length and width show good agreement with the reconnection
timescale and anastrophy-conservation scaling, $t^{1/2}$. In
the helical case, there is only approximate agreement with the predicted
scaling, $t^{4/7}$. Nonetheless, it is interesting and reaffirming that the
evolution of length scales obtained from this local analysis is broadly
consistent with the theoretical SP predicted scalings.

\bibliography{refpaper}{}
\bibliographystyle{aasjournalv7}




\end{document}